\documentclass[12pt]{article}

\usepackage[margin=1in]{geometry}

\usepackage[utf8]{inputenc}
\usepackage[T1]{fontenc}
\usepackage{lmodern}

\usepackage[english]{babel}

\usepackage{setspace}
\usepackage{parskip}

\usepackage{graphicx}
\usepackage{graphics}
\usepackage{float}
\usepackage{placeins}
\usepackage{pdflscape}
\usepackage{pdfpages}

\usepackage{booktabs}
\usepackage{array}
\usepackage{tabularx}
\usepackage{multirow}
\usepackage{longtable}
\usepackage{makecell}

\usepackage[labelsep=period,font=sc,skip=0pt,justification=centering]{caption}
\usepackage{subcaption}
\usepackage{titling}
\usepackage{amsmath,amssymb,amsthm}
\usepackage{mathtools}
\usepackage{amscd}
\usepackage{amsfonts}
\usepackage{amsbsy}
\usepackage{bm}
\usepackage{dsfont}
\usepackage{cancel}
\usepackage{mathrsfs}
\usepackage{latexsym}
\usepackage{braket}

\usepackage[shortlabels]{enumitem}

\usepackage{booktabs}
\usepackage{tikz}
\usetikzlibrary{
    arrows.meta,
    positioning,
    backgrounds,
    calc,
    fit,
    shadows,
    shapes.geometric,
    decorations.pathreplacing,
}

\usepackage{pifont}

\usepackage{xcolor}
\definecolor{AcceptGreen}{RGB}{46,139,87}
\definecolor{WarningOrange}{RGB}{230,126,34}
\definecolor{DangerRed}{RGB}{192,57,43}
\definecolor{CardinalRed}{cmyk}{0,1,0.65,0.34}

\definecolor{scblue}{HTML}{2563EB}        
\definecolor{scbluedark}{HTML}{1E40AF}    
\definecolor{scbluelight}{HTML}{DBEAFE}   
\definecolor{scbluemid}{HTML}{93C5FD}     
\definecolor{scbluepale}{HTML}{EFF6FF}    
\definecolor{scgray}{HTML}{475569}        
\definecolor{scgraybg}{HTML}{F8FAFC}      
\definecolor{termbg}{HTML}{1E293B}        
\definecolor{termfg}{HTML}{E2E8F0}        
\definecolor{termaccent}{HTML}{93C5FD}    
\definecolor{agleader}{HTML}{334155}       
\definecolor{agleaderbg}{HTML}{E2E8F0}
\definecolor{agplanner}{HTML}{1E40AF}      
\definecolor{agplannerbg}{HTML}{DBEAFE}
\definecolor{agbuilder}{HTML}{166534}      
\definecolor{agbuilderbg}{HTML}{DCFCE7}
\definecolor{agtester}{HTML}{9A3412}       
\definecolor{agtesterbg}{HTML}{FEF3C7}
\definecolor{agsimulator}{HTML}{6B21A8}    
\definecolor{agsimulatorbg}{HTML}{F3E8FF}
\definecolor{agscriber}{HTML}{115E59}      
\definecolor{agscriberbg}{HTML}{CCFBF1}
\definecolor{agreviewer}{HTML}{991B1B}     
\definecolor{agreviewerbg}{HTML}{FEE2E2}
\definecolor{agshipper}{HTML}{78350F}      
\definecolor{agshipperbg}{HTML}{FEF3C7}
\definecolor{controlblue}{HTML}{2563EB}
\definecolor{bridgepurple}{HTML}{7C3AED}
\definecolor{codeorange}{HTML}{EA580C}
\definecolor{testgreen}{HTML}{059669}
\definecolor{simteal}{HTML}{0D9488}
\definecolor{recordamber}{HTML}{D97706}
\definecolor{convergered}{HTML}{DC2626}
\definecolor{shipgray}{HTML}{475569}
\definecolor{borderlight}{HTML}{E2E8F0}
\definecolor{stateblue}{HTML}{DBEAFE}
\definecolor{signalred}{HTML}{FEE2E2}
\definecolor{signalamber}{HTML}{FEF3C7}
\definecolor{signalgreen}{HTML}{D1FAE5}
\definecolor{artifactfill}{HTML}{F5F3FF}
\definecolor{sigpass}{HTML}{166534}        
\definecolor{sigpassbg}{HTML}{DCFCE7}
\definecolor{sigpassborder}{HTML}{86EFAC}
\definecolor{sigblock}{HTML}{991B1B}       
\definecolor{sigblockbg}{HTML}{FEE2E2}
\definecolor{sigblockborder}{HTML}{FCA5A5}

\definecolor{linkblue}{HTML}{2563EB}

\usepackage[T1]{fontenc}
\usepackage{lmodern}
\usepackage{amssymb}
\usepackage{tikz}
\usetikzlibrary{arrows.meta, positioning, shapes.geometric, calc,
                 decorations.pathreplacing, shadows, backgrounds}
\usepackage{xcolor}

\definecolor{controlblue}{HTML}{2563EB}
\definecolor{bridgepurple}{HTML}{7C3AED}
\definecolor{codeorange}{HTML}{EA580C}
\definecolor{testgreen}{HTML}{059669}
\definecolor{simteal}{HTML}{0D9488}
\definecolor{recordamber}{HTML}{D97706}
\definecolor{convergered}{HTML}{DC2626}
\definecolor{shipgray}{HTML}{475569}
\definecolor{borderlight}{HTML}{E2E8F0}
\definecolor{artifactfill}{HTML}{F5F3FF}
\definecolor{holdamber}{HTML}{B45309}
\definecolor{blockred}{HTML}{DC2626}
\definecolor{stoppurple}{HTML}{7C3AED}
\definecolor{isowall}{HTML}{DC2626}
\definecolor{signalbg}{HTML}{FFF8F0}

\usepackage[colorlinks,linkcolor=linkblue,citecolor=linkblue,urlcolor=linkblue]{hyperref}
\usepackage{pdfpages}
\usepackage{url}

\usepackage[most]{tcolorbox}
\tcbuselibrary{listings,skins,breakable}

\newtcolorbox{userprompt}[1][]{%
  enhanced,
  breakable,
  colback=scbluepale,
  colframe=scbluemid,
  fonttitle=\sffamily\bfseries\small,
  title={\raisebox{-1pt}{\footnotesize\texttt{\$>}}~User Prompt},
  coltitle=scbluedark,
  boxrule=0.8pt,
  arc=4pt,
  left=8pt, right=8pt, top=4pt, bottom=4pt,
  fontupper=\itshape,
  #1
}

\newtcolorbox{artifact}[1][]{%
  enhanced,
  breakable,
  colback=scbluepale,
  colframe=scbluemid!70,
  fonttitle=\sffamily\bfseries\small\ttfamily,
  coltitle=scbluedark,
  boxrule=0.6pt,
  arc=3pt,
  left=8pt, right=8pt, top=4pt, bottom=4pt,
  fontupper=\small,
  #1
}

\newtcolorbox{terminal}[1][]{%
  enhanced,
  breakable,
  colback=termbg,
  colframe=termbg,
  colupper=termfg,
  fontupper=\ttfamily\footnotesize,
  arc=4pt,
  left=10pt, right=8pt, top=6pt, bottom=6pt,
  #1
}

\newtcolorbox{holdsignal}[1][]{%
  enhanced,
  colback=scbluepale,
  colframe=scbluemid,
  fonttitle=\sffamily\bfseries\small,
  title={\textsc{Hold}},
  coltitle=scbluedark,
  boxrule=0.8pt,
  arc=3pt,
  left=8pt, right=8pt, top=3pt, bottom=3pt,
  fontupper=\small,
  #1
}

\newtcolorbox{blocksignal}[1][]{%
  enhanced,
  colback=sigblockbg,
  colframe=sigblockborder,
  fonttitle=\sffamily\bfseries\small,
  title={\textcolor{sigblock}{\ding{55}}~\textsc{Block}},
  coltitle=sigblock,
  boxrule=0.8pt,
  arc=3pt,
  left=8pt, right=8pt, top=3pt, bottom=3pt,
  fontupper=\small,
  #1
}

\newtcolorbox{passsignal}[1][]{%
  enhanced,
  colback=sigpassbg,
  colframe=sigpassborder,
  fonttitle=\sffamily\bfseries\small,
  title={\textcolor{sigpass}{\ding{51}}~\textsc{Pass}},
  coltitle=sigpass,
  boxrule=0.8pt,
  arc=3pt,
  left=8pt, right=8pt, top=3pt, bottom=3pt,
  fontupper=\small,
  #1
}

\newcommand{\agentlabel}[1]{\textsc{\textbf{#1}}}
\newcommand{\aleader}{\agentlabel{Leader}}
\newcommand{\aplanner}{\agentlabel{Planner}}
\newcommand{\abuilder}{\agentlabel{Builder}}
\newcommand{\atester}{\agentlabel{Tester}}
\newcommand{\asimulator}{\agentlabel{Simulator}}
\newcommand{\ascriber}{\agentlabel{Scriber}}
\newcommand{\areviewer}{\agentlabel{Reviewer}}
\newcommand{\ashipper}{\agentlabel{Shipper}}

\usepackage{listings}
\lstdefinelanguage{Stata}{
  morekeywords={if, first, cluster},
  sensitive=true,
}

\usepackage[flushmargin]{footmisc}
\usepackage[round]{natbib}

\usepackage{xargs}
\usepackage{lineno}
\usepackage{CJKutf8} 
\usepackage{epsfig}  

\usepackage{titlesec}
\titlelabel{\thetitle.\ }
\titleformat{\section}
{\normalfont\fontsize{15}{15}\bfseries}{\thesection.}{0.5em}{}

\usepackage{todonotes}
\makeatletter
\providecommand\@dotsep{5}
\def\listtodoname{List of Todos}
\def\listoftodos{\@starttoc{tdo}\listtodoname}
\makeatother

\makeatletter
\def\maxwidth{\ifdim\Gin@nat@width>\linewidth\linewidth\else\Gin@nat@width\fi}
\def\maxheight{\ifdim\Gin@nat@height>\textheight\textheight\else\Gin@nat@height\fi}
\makeatother
\setkeys{Gin}{width=\maxwidth,height=\maxheight,keepaspectratio}

\graphicspath{{./graphs/}}%

\newcommand{\bland}{\begin{landscape}}
\newcommand{\eland}{\end{landscape}}

\newcommand{\burl}[1]{\textcolor{blue}{\url{#1}}}

\newenvironment{itemize*}%
  {\begin{itemize}\setlength{\itemsep}{0pt}\setlength{\parskip}{0pt}}%
  {\end{itemize}}
\newenvironment{enumerate*}%
  {\begin{enumerate}\setlength{\itemsep}{0pt}\setlength{\parskip}{0pt}}%
  {\end{enumerate}}

\newcommand{\beq}{\begin{equation}}
\newcommand{\eeq}{\end{equation}}

\newcommand*\Bigpar[1]{\left( #1 \right)}

\newcommandx{\deriv}[2][1=x,2=f]{\nabla \, #2 \Bigpar{ #1 } }

\renewcommand{\to}{\rightarrow}

\newtheoremstyle{mystyle}
  {12pt}{12pt}{}{}{\sffamily\bfseries}{.}{0.5em}
  {\thmname{#1}\thmnumber{ #2}\thmnote{ (#3)}}
\theoremstyle{mystyle}

\newenvironment{proof-sketch}{\noindent{\bf Sketch of Proof}\hspace*{1em}}{\qed\bigskip\\}
\newenvironment{proof-idea}{\noindent{\bf Proof Idea}\hspace*{1em}}{\qed\bigskip\\}
\newenvironment{proof-of-lemma}[1][{}]{\noindent{\bf Proof of Lemma {#1}}\hspace*{1em}}{\qed\bigskip\\}
\newenvironment{proof-of-proposition}[1][{}]{\noindent{\bf Proof of Proposition {#1}}\hspace*{1em}}{\qed\bigskip\\}
\newenvironment{proof-of-theorem}[1][{}]{\noindent{\bf Proof of Theorem {#1}}\hspace*{1em}}{\qed\bigskip\\}
\newenvironment{inner-proof}{\noindent{\bf Proof}\hspace{1em}}{$\bigtriangledown$\medskip\\}
\newenvironment{proof-attempt}{\noindent{\bf Proof Attempt}\hspace*{1em}}{\qed\bigskip\\}

\newcolumntype{L}[1]{>{\raggedright\let\newline\\\arraybackslash\hspace{0pt}}m{#1}}
\newcolumntype{C}[1]{>{\centering\let\newline\\\arraybackslash\hspace{0pt}}m{#1}}
\newcolumntype{R}[1]{>{\raggedleft\let\newline\\\arraybackslash\hspace{0pt}}m{#1}}

\graphicspath{{./figs/}}%

\begin{document}


\title{\Large\bf StatsClaw: An AI-Collaborative Workflow\\for Statistical Software Development\thanks{Tianzhu Qin, PhD Candidate, The Centre for Human-Inspired Artificial Intelligence, University of Cambridge. Email: \url{tq224@cam.ac.uk}. Yiqing Xu, Assistant Professor, Department of Political Science, Stanford University. Email: \url{yiqingxu@stanford.edu}. The authors used Claude Code as research, coding, and writing assistants in preparing this manuscript. All interpretations, conclusions, and any errors remain solely the responsibility of the authors. We thank Euphie Wang for designing the StatsClaw mascot.}
\\\raisebox{-2pt}{\includegraphics[height=3em]{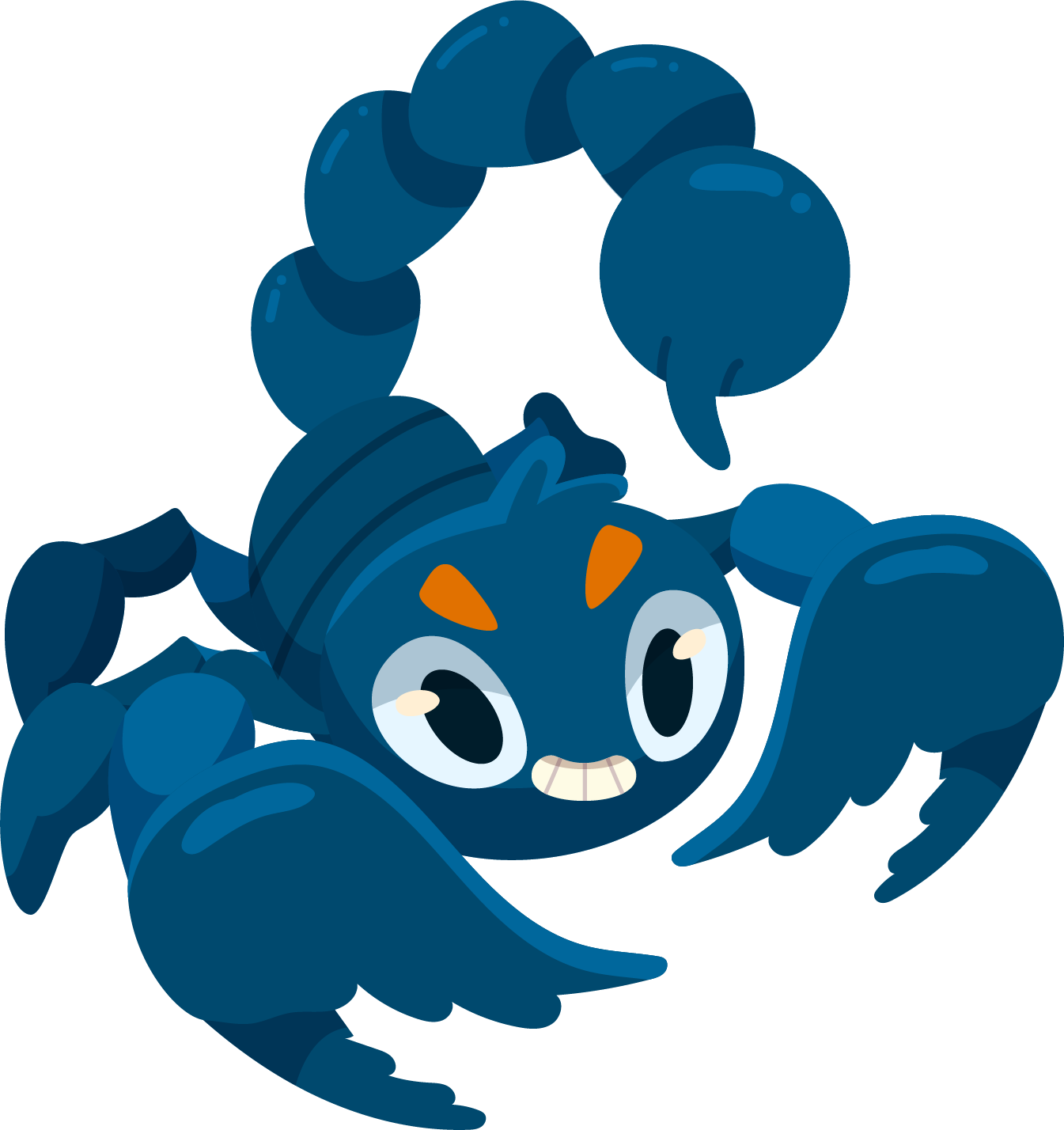}}%
\\\bigskip}
\vspace{1.5em}

\author{
Tianzhu Qin\\(Cambridge)
\and
Yiqing Xu\\(Stanford)
}

\date{\bigskip April 4, 2026}

\maketitle
\vspace{1em}
\begin{abstract}
\vspace{1em}
\noindent Translating statistical methods into reliable software is a persistent bottleneck in quantitative research. Existing AI code-generation tools produce code quickly but cannot guarantee faithful implementation---a critical requirement for statistical software. We introduce \textsc{StatsClaw}, a multi-agent architecture for Claude Code that enforces information barriers between code generation and validation. A planning agent produces independent specifications for implementation, simulation, and testing, dispatching them to separate agents that cannot see each other's instructions: the builder implements without knowing the ground-truth parameters, the simulator generates data without knowing the algorithm, and the tester validates using deterministic criteria. We describe the approach, demonstrate it end-to-end on a probit estimation package, and evaluate it across three applications to the authors' own R and Python packages. The results show that structured AI-assisted workflows can absorb the engineering overhead of the software lifecycle while preserving researcher control over every substantive methodological decision.

\end{abstract}

\thispagestyle{empty}
\clearpage
\newpage
\doublespace

\clearpage
\setcounter{page}{1}
\doublespacing

\section{Introduction}
\label{sec:intro}

Methodological innovation in statistics and the social sciences begins with theoretical analysis but ultimately depends on software. A new estimator reaches practitioners through a package---an R library on CRAN, a Python module on PyPI, a Stata command on SSC. Producing that package requires translating mathematical derivations into algorithms, implementing those algorithms in code, writing tests that verify numerical correctness, building documentation that explains usage, and maintaining all of these as the package evolves. For most methodological teams, this engineering work is the binding constraint \citep{ram2019community}.

The scale of the problem has grown. Modern causal inference packages support multiple estimators, cross-validation procedures, bootstrap inference, C++ backends, and visualization systems. Keeping code, documentation, and tests mutually consistent across these components is labor-intensive and error-prone. When a research team modifies a scoring function, the change propagates through parameter interfaces, bootstrap routines, plotting code, and user-facing documentation. Missing a single downstream dependency produces bugs that may be silent---numerically plausible but statistically incorrect \citep{mccullough1999numerical}.

These costs have consequences. Methods that lack reliable software are adopted slowly, if at all. Packages that ship without adequate testing may produce incorrect results in edge cases. Documentation that falls out of sync with code misleads users about what the software actually computes. The result is a gap between the methods that exist in the literature and the methods that are available in practice \citep{peng2011reproducible}.

Recent advances in AI code generation have begun to narrow this gap. Large language models can now produce syntactically valid code at impressive speed, from function-level generation \citep{chen2021evaluating} to repository-level issue resolution \citep{jimenez2024swebench} and agentic interaction with codebases \citep{yang2024sweagent}. But for statistical software, speed of generation is not the binding constraint---correctness is. The standard workflow for AI-assisted development is \textit{generate, then test}: the same model (or the same information set) produces both the code and the tests. This creates a structural problem. If the generator misunderstands the specification---conflating the observed and expected information matrix, using the wrong covariance estimator variant, mishandling truncation in MCMC sampling---it will likely embed the same misunderstanding in its tests. The code passes, but the implementation is wrong.

This paper introduces \textsc{StatsClaw}, an AI-assisted workflow designed around a simple principle: \textit{the process that generates code must never be the same process that validates it}. Concretely, \textsc{StatsClaw} is a set of agent prompts and configuration files for Anthropic's Claude Code---there is nothing to install beyond Claude Code itself, and the ``agents'' are dispatched within a single session rather than running as separate processes. The framework's value lies entirely in the \textit{structure} it imposes: information barriers between code generation and code validation, mandatory comprehension checks before any code is written, and a state machine that enforces sequential gates. A planning agent reads the source material (mathematical derivations, existing codebases, pseudocode, algorithm descriptions) and produces two independent documents: a \textit{specification} telling a builder agent what to implement, and a \textit{test specification} telling a tester agent what to verify. Neither agent sees the other's document. When the workflow includes simulation-based diagnostics, a third document drives an independent Monte Carlo evaluation that treats the implementation as a black box.

We describe the approach (Section~\ref{sec:approach}), walk through a complete example from source material to installable R package (Section~\ref{sec:tutorial}), and evaluate the workflow across three applications of increasing complexity (Section~\ref{sec:applications}): paper-to-feature development, code translation, and sustained multi-day refactoring. Section~\ref{sec:discussion} discusses limitations, scope, and implications. Appendix~\ref{sec:appendix} provides the full architecture reference and a practical guide for adoption.

\section{The Approach}
\label{sec:approach}

The central problem in AI-assisted statistical software development is not code generation but \textit{verification}. A language model that misunderstands a mathematical specification will produce code that is syntactically valid, passes superficial checks, and silently computes the wrong thing. The standard generate-then-test workflow amplifies this risk: the same information that produced the misunderstanding also produces the tests, so errors correlate rather than cancel. \textsc{StatsClaw} addresses this by structuring AI-assisted development around specialized agents, enforced information barriers, and mandatory comprehension checks.

\subsection{Agents and Orchestration}
\label{sec:approach-agents}

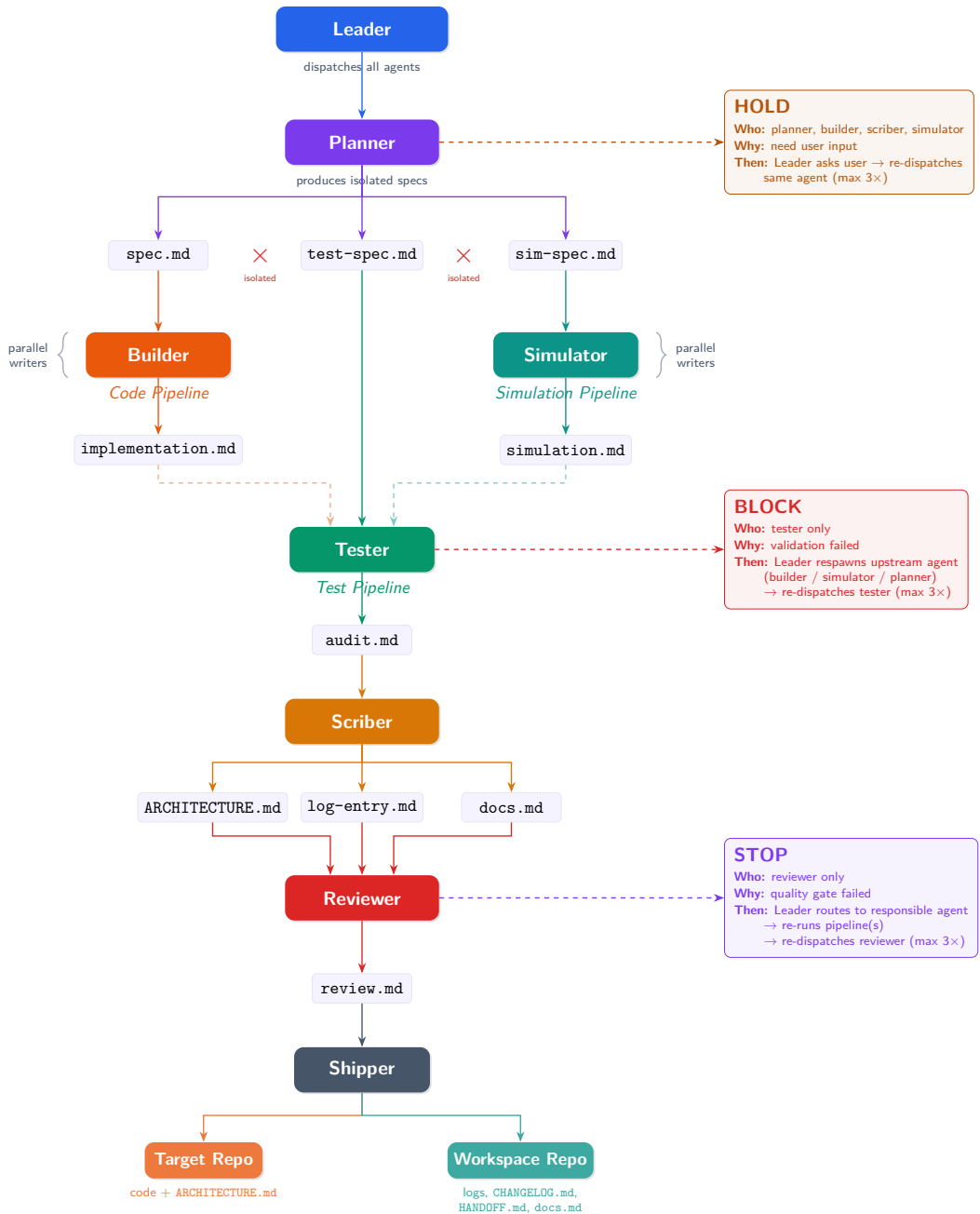
\begin{figure}[!ht]
\centering
\resizebox{0.85\textwidth}{!}{\begin{tikzpicture}[
    >=Stealth,
    agent/.style={
        rectangle, rounded corners=6pt, minimum width=3.2cm, minimum height=1cm,
        font=\sffamily\bfseries, text=white,
        drop shadow={shadow xshift=1pt, shadow yshift=-1pt, opacity=0.25}
    },
    artifact/.style={
        rectangle, rounded corners=3pt, minimum width=2.2cm, minimum height=0.65cm,
        font=\sffamily\small, fill=artifactfill, draw=borderlight, text=black
    },
    signal/.style={
        rectangle, rounded corners=4pt, minimum width=3.8cm, minimum height=1.2cm,
        draw=#1, fill=#1!6, text=#1, font=\sffamily\scriptsize,
        align=left, inner sep=6pt
    },
    flowline/.style={->, thick, color=#1},
    dashedflow/.style={->, dashed, color=#1, thick},
    note/.style={font=\sffamily\scriptsize, text=shipgray, align=center},
    pipelabel/.style={font=\sffamily\small\itshape, text=#1},
]


\node[agent, fill=controlblue, minimum width=3.8cm] (leader) at (0, 0) {Leader};
\node[note, below=0.08cm of leader] {dispatches all agents};

\node[agent, fill=bridgepurple, minimum width=3.4cm] (planner) at (0, -2.5) {Planner};
\node[note, below=0.08cm of planner] {produces isolated specs};

\draw[flowline=controlblue] (leader.south) -- (planner.north);

\node[artifact] (specmd) at (-4.5, -5) {\texttt{spec.md}};
\node[artifact] (testspecmd) at (0, -5) {\texttt{test-spec.md}};
\node[artifact] (simspecmd) at (4.5, -5) {\texttt{sim-spec.md}};

\draw[flowline=bridgepurple] (planner.south) -- ++(0,-0.7) -| (specmd.north);
\draw[flowline=bridgepurple] (planner.south) -- (testspecmd.north);
\draw[flowline=bridgepurple] (planner.south) -- ++(0,-0.7) -| (simspecmd.north);

\node[font=\Large\bfseries, text=isowall] at (-2.25, -5) {$\times$};
\node[font=\Large\bfseries, text=isowall] at (2.25, -5) {$\times$};
\node[note, text=isowall, font=\sffamily\tiny] at (-2.25, -5.5) {isolated};
\node[note, text=isowall, font=\sffamily\tiny] at (2.25, -5.5) {isolated};

\node[agent, fill=codeorange] (builder) at (-4.5, -7.2) {Builder};
\node[pipelabel=codeorange, below=0.05cm of builder] {Code Pipeline};

\node[agent, fill=simteal] (simulator) at (4.5, -7.2) {Simulator};
\node[pipelabel=simteal, below=0.05cm of simulator] {Simulation Pipeline};

\draw[flowline=codeorange] (specmd.south) -- (builder.north);
\draw[flowline=simteal] (simspecmd.south) -- (simulator.north);

\draw[decorate, decoration={brace, amplitude=6pt}, thick, shipgray!50]
    (-6.5, -7.7) -- (-6.5, -6.7)
    node[midway, left=8pt, note] {parallel\\writers};
\draw[decorate, decoration={brace, amplitude=6pt, mirror}, thick, shipgray!50]
    (6.5, -7.7) -- (6.5, -6.7)
    node[midway, right=8pt, note] {parallel\\writers};

\node[artifact] (implmd) at (-4.5, -9.3) {\texttt{implementation.md}};
\node[artifact] (simmd) at (4.5, -9.3) {\texttt{simulation.md}};

\draw[flowline=codeorange] (builder.south) -- (implmd.north);
\draw[flowline=simteal] (simulator.south) -- (simmd.north);

\node[agent, fill=testgreen] (tester) at (0, -11.5) {Tester};
\node[pipelabel=testgreen, below=0.05cm of tester] {Test Pipeline};

\draw[flowline=testgreen] (testspecmd.south) -- (tester.north);

\draw[dashedflow=codeorange!50]
    (implmd.south) -- ++(0,-0.4) -| ([xshift=-0.7cm]tester.north);
\draw[dashedflow=simteal!50]
    (simmd.south) -- ++(0,-0.4) -| ([xshift=0.7cm]tester.north);

\node[artifact] (auditmd) at (0, -13.5) {\texttt{audit.md}};
\draw[flowline=testgreen] (tester.south) -- (auditmd.north);

\node[agent, fill=recordamber, minimum width=3.4cm] (scriber) at (0, -15.3) {Scriber};

\draw[flowline=recordamber] (auditmd.south) -- (scriber.north);

\node[artifact] (archmd) at (-3.3, -17.2) {\texttt{ARCHITECTURE.md}};
\node[artifact] (logmd) at (0, -17.2) {\texttt{log-entry.md}};
\node[artifact] (docsmd) at (3.3, -17.2) {\texttt{docs.md}};

\draw[flowline=recordamber] (scriber.south) -- ++(0,-0.4) -| (archmd.north);
\draw[flowline=recordamber] (scriber.south) -- (logmd.north);
\draw[flowline=recordamber] (scriber.south) -- ++(0,-0.4) -| (docsmd.north);

\node[agent, fill=convergered, minimum width=3.4cm] (reviewer) at (0, -19.2) {Reviewer};

\draw[flowline=convergered] (archmd.south) -- ++(0,-0.3) -| ([xshift=-0.7cm]reviewer.north);
\draw[flowline=convergered] (logmd.south) -- (reviewer.north);
\draw[flowline=convergered] (docsmd.south) -- ++(0,-0.3) -| ([xshift=0.7cm]reviewer.north);

\node[artifact] (reviewmd) at (0, -21.2) {\texttt{review.md}};
\draw[flowline=convergered] (reviewer.south) -- (reviewmd.north);

\node[agent, fill=shipgray, minimum width=3cm] (shipper) at (0, -23) {Shipper};

\draw[flowline=shipgray] (reviewmd.south) -- (shipper.north);

\node[agent, fill=codeorange!80, minimum width=2.6cm, minimum height=0.8cm,
      font=\sffamily\small\bfseries] (targetrepo) at (-3.5, -25) {Target Repo};
\node[agent, fill=simteal!80, minimum width=2.6cm, minimum height=0.8cm,
      font=\sffamily\small\bfseries] (workspacerepo) at (3.5, -25) {Workspace Repo};

\draw[flowline=codeorange!80] (shipper.south) -- ++(0,-0.5) -| (targetrepo.north);
\draw[flowline=simteal!80] (shipper.south) -- ++(0,-0.5) -| (workspacerepo.north);

\node[note, text=codeorange!80, below=0.05cm of targetrepo]
    {code + \texttt{ARCHITECTURE.md}};
\node[note, text=simteal!80, below=0.05cm of workspacerepo]
    {logs, \texttt{CHANGELOG.md},\\\texttt{HANDOFF.md}, \texttt{docs.md}};


\node[signal=holdamber, anchor=west] (holdbox) at (8, -2.5) {%
    \textbf{\normalsize HOLD}\\[3pt]
    \textbf{Who:} planner, builder, scriber, simulator\\[1pt]
    \textbf{Why:} need user input\\[1pt]
    \textbf{Then:} Leader asks user $\to$ re-dispatches\\
    \hspace{2.2em}same agent (max 3$\times$)
};
\draw[dashedflow=holdamber] (planner.east) -- ++(1,0) -- (holdbox.west);

\node[signal=blockred, anchor=west] (blockbox) at (8, -11.5) {%
    \textbf{\normalsize BLOCK}\\[3pt]
    \textbf{Who:} tester only\\[1pt]
    \textbf{Why:} validation failed\\[1pt]
    \textbf{Then:} Leader respawns upstream agent\\
    \hspace{2.2em}(builder / simulator / planner)\\
    \hspace{2.2em}$\to$ re-dispatches tester (max 3$\times$)
};
\draw[dashedflow=blockred] (tester.east) -- ++(1,0) -- (blockbox.west);

\node[signal=stoppurple, anchor=west] (stopbox) at (8, -19.2) {%
    \textbf{\normalsize STOP}\\[3pt]
    \textbf{Who:} reviewer only\\[1pt]
    \textbf{Why:} quality gate failed\\[1pt]
    \textbf{Then:} Leader routes to responsible agent\\
    \hspace{2.2em}$\to$ re-runs pipeline(s)\\
    \hspace{2.2em}$\to$ re-dispatches reviewer (max 3$\times$)
};
\draw[dashedflow=stoppurple] (reviewer.east) -- ++(1,0) -- (stopbox.west);

\end{tikzpicture}}
\caption{\textsc{StatsClaw} workflow architecture. The planner produces three isolated specification documents; the builder, tester, and simulator each receive only their own specification ($\times$ marks information barriers). The reviewer cross-compares all pipeline outputs before issuing a ship verdict.}\label{fig:architecture}
\end{figure}

\textsc{StatsClaw} orchestrates eight specialized agents\footnote{The current implementation includes a ninth agent, the \emph{distiller}, which supports an optional shared-knowledge system (``Brain mode'') currently under development. The distiller extracts reusable, privacy-scrubbed insights from completed workflow runs and, with explicit user consent, contributes them to a shared knowledge repository. Brain mode is planned for a future release and is not discussed in this paper. When it is disabled---the default---the eight-agent architecture described here is complete.} within a single Claude Code session (Figure~\ref{fig:architecture}). Each agent has a defined role, a restricted information set, and a fixed position in the workflow's state machine. The system is implemented as a set of agent prompts and configuration files for Claude Code; the ``agents'' are dispatched via Claude Code's built-in Agent tool within a single session, not as separate processes or services.

\aleader{} is the orchestrator. It receives the user's request, selects the appropriate workflow type (from ten supported patterns), dispatches all other agents, and enforces a state machine with mandatory preconditions at each transition. It is the only agent that communicates directly with the user. Not every task requires the full pipeline: a simple bug fix may need only the builder and tester, while a greenfield package invokes the planner, all three execution agents, and the full review chain. The leader makes this decision, which is why it is a role distinct from the planner. \aplanner{} is the bridge between the user's source material and the execution pipelines. It reads all inputs---mathematical derivations, existing codebases, pseudocode, algorithm descriptions---and produces the specification documents that downstream agents will consume. The planner is the only agent with access to the full information set; every other agent sees only the slice it needs.

Three agents form the execution layer, each operating on its own specification in an isolated git worktree. \abuilder{} implements the software: writing source code, configuring build systems, and producing an installable package. \asimulator{} designs and runs simulation experiments that treat the implementation as a black box. \atester{} independently constructs a validation suite: unit tests, edge cases, property-based checks, and cross-reference benchmarks against established implementations. The tester's acceptance criteria are hard-coded and version-controlled: matching ground-truth parameters within a specified tolerance is a deterministic check, not a judgment call by the language model. These three agents execute in parallel and cannot communicate with one another. 

\ascriber{} documents the process after the execution pipelines complete, recording architecture decisions, implementation details, and a structured log of what each agent did. Because the scriber operates with a long accumulated context, it risks \textit{rule drift}---forgetting constraints imposed earlier in the session. \areviewer{} mitigates this by receiving only the scriber's condensed output, enforcing rules from a fresh context window. The reviewer is the convergence gate: it is the only agent besides the planner with cross-pipeline visibility, reading all specification documents, all pipeline outputs, and the scriber's documentation, then auditing whether the pipelines converge---whether the builder's implementation satisfies the tester's behavioral contracts and the simulator's Monte Carlo criteria. The reviewer issues a ship or no-ship verdict. Finally, \ashipper{} handles git operations (commit, push, pull request) upon user authorization.

Three interrupt signals coordinate the workflow when things go wrong. A \textsc{Hold} signal, which any agent can raise, requests user input; the leader forwards the question and re-dispatches upon receiving an answer. A \textsc{Block} signal, which only the tester can raise, indicates a validation failure; the leader re-dispatches the builder with the failure details, and the tester re-validates after the fix. A \textsc{Stop} signal, which only the reviewer can raise, indicates a quality-gate failure and routes back to the responsible agent. Each signal permits a maximum of three retry cycles before escalating to the user. The state machine progresses through nine states---from credential verification through planning, specification, pipeline execution, documentation, review, and shipping---with hard gates at each transition. A detailed description of each agent's responsibilities, the full state machine, and the ten supported workflow types is provided in Appendix~\ref{sec:appendix}.

\subsection{Information Barriers}
\label{sec:approach-barriers}

The architecture described above would offer little advantage over a single-agent workflow if all agents shared the same information. The key design constraint is what each agent \textit{cannot} see.

In the full workflow, the planner produces three self-contained specification documents from the source material. The implementation specification (\texttt{spec.md}) describes the algorithm's computational steps, data structures, input validation rules, and API contracts; it is given to the builder. The simulation specification (\texttt{sim-spec.md}) specifies the data-generating process, scenario grid, performance metrics, and acceptance criteria; it is given to the simulator. The test specification (\texttt{test-spec.md}) describes expected behaviors, numerical tolerances, edge cases, and property-based invariants; it is given to the tester. Each document is independently sufficient for its recipient. Neither references the other. At dispatch time, the builder receives only the implementation specification and never sees the test specification; the tester receives only the test specification and never sees the implementation specification or the source code. Table~\ref{tab:isolation} summarizes the full access matrix.

\begin{table}[ht]
\centering
\small
\caption{Information barriers: each agent sees only its own specification.}\label{tab:isolation}
\begin{tabular}{lccc}
\toprule
\textbf{Agent} & Model Spec. & Simulation Spec. & Test Spec. \\
\midrule
\abuilder{}    & \textcolor{sigpass}{\checkmark} & \textcolor{sigblock}{$\times$} & \textcolor{sigblock}{$\times$} \\[3pt]
\asimulator{}  & \textcolor{sigblock}{$\times$} & \textcolor{sigpass}{\checkmark} & \textcolor{sigblock}{$\times$} \\[3pt]
\atester{}     & \textcolor{sigblock}{$\times$} & \textcolor{sigblock}{$\times$} & \textcolor{sigpass}{\checkmark} \\[3pt]
\bottomrule
\end{tabular}
\end{table}

This separation prevents each agent from \textit{teaching to the test}---finding surface-level solutions that pass validation without genuinely satisfying the specification. The failure modes are concrete and predictable. If the builder knows the ground-truth parameters used for validation, it can hardcode those values rather than implementing the algorithm correctly---the tests pass, but the implementation is wrong. If the simulator knows how the algorithm works, it can design trivially easy data-generating processes that the implementation passes without being tested on difficult cases. If the tester sees the source code, it may design tests that follow the implementation's logic rather than independently verifying behavioral contracts. By enforcing information barriers, \textsc{StatsClaw} ensures that the builder must implement the algorithm from its specification alone, the simulator must design DGPs from the mathematical theory alone, and the tester must validate purely by checking whether ground truth is recovered---a deterministic comparison against the planner's criteria. A bug that survives must simultaneously satisfy two or three independently derived behavioral contracts, analogous to independent replication in experimental science.

\subsection{Deep Comprehension as a Hard Gate}
\label{sec:approach-comprehension}

Information barriers ensure that downstream agents cannot contaminate each other's work, but they do nothing to prevent the planner itself from misunderstanding the source material. If the planner misreads a formula or conflates two estimator variants, it will embed that misunderstanding in all three specification documents, and the information barriers will faithfully propagate a coherent but wrong set of instructions. \textsc{StatsClaw} addresses this through a mandatory comprehension protocol that serves as a hard gate before any specification is written.

The protocol requires the planner to inventory every equation, symbol, assumption, and theorem from the source material; restate all mathematical content in its own notation, defining every symbol's type, dimensions, and domain; self-test against diagnostic questions (can the core requirement be restated without reference to the source? can every formula be reproduced and explained? have implicit assumptions been identified?); and produce an auditable \texttt{comprehension.md} artifact that the user can review before any code is generated. This artifact is the single point of human oversight in the workflow: if the planner's understanding is correct, the downstream specifications will be correct; if it is wrong, the user catches it here rather than debugging silent numerical errors downstream.

If the protocol yields only partial understanding, the agent raises a \textsc{Hold} signal---a structured request for user clarification. The system does not proceed until comprehension is verified. This shifts the failure mode from ``generated wrong code'' to ``asked the right clarifying questions.''

\section{Demonstration: Implementing Probit Regressions}
\label{sec:tutorial}

To make the workflow concrete, we walk through a complete application: building an R package implementing three probit estimation methods from a short PDF specification, with Monte Carlo simulation to evaluate finite-sample performance. This example uses the three-pipeline architecture (builder, tester, and simulator) and illustrates every stage of the workflow.

\subsection{The Problem}
\label{sec:tutorial-problem}

The probit model is a standard binary choice model. The outcome is generated through a latent variable $y_i^* = \mathbf{x}_i'\boldsymbol{\beta} + \varepsilon_i$, where $\varepsilon_i \sim \mathcal{N}(0,1)$, and the observed binary outcome is $y_i = \mathds{1}(y_i^* > 0)$. The source material is a 4-page PDF (reproduced in Appendix~\ref{sec:probit-spec}) specifying three estimation methods for this model: (a)~\textit{maximum likelihood via Newton-Raphson}, which directly maximizes the log-likelihood using its gradient and Hessian, with the inverse Mills ratio ($\phi/\Phi$) for numerical stability; (b)~a \textit{Bayesian Gibbs sampler} based on the Albert-Chib data augmentation scheme \citep{albert1993bayesian}, which augments the model with latent $y_i^*$ drawn from a truncated normal and then samples $\boldsymbol{\beta}$ from the resulting conjugate normal posterior; and (c)~\textit{random-walk Metropolis-Hastings}, which proposes $\boldsymbol{\beta}$ from a random walk and accepts or rejects via the likelihood ratio.

All three methods target the same parameter $\boldsymbol{\beta}$, so their estimates should agree asymptotically, making correct implementation verifiable through Monte Carlo comparison. Probit estimation is also a solved problem in R: \texttt{glm(family = binomial(link = "probit"))} provides a trusted reference implementation. \textsc{StatsClaw} is configured to implement algorithms from scratch rather than wrap existing packages, so the \texttt{glm()} output serves as an independent benchmark against which the builder's implementation can be validated.

\subsection{Setup and Prompt}
\label{sec:tutorial-setup}

The user opens Claude Code with the \textsc{StatsClaw} configuration loaded and provides a single prompt:

\begin{userprompt}
Build the R works from this PDF. Three probit estimation methods in C++ via Rcpp/Armadillo: MLE (Newton-Raphson), Bayesian Gibbs sampler (Albert-Chib data augmentation), and random-walk Metropolis-Hastings. After building, run a Monte Carlo simulation comparing all three on bias, RMSE, coverage, and computational speed across $N = \{200, 500, 1000, 5000\}$ with 500 replications per scenario. Ship it.
\end{userprompt}

\aleader{} resolves the target repository, detects no existing language profile, and selects the full workflow, which activates all agents including \abuilder, \asimulator, and \atester.

\subsection{Comprehension}
\label{sec:tutorial-comprehension}

\aplanner{} ingests the PDF and executes the comprehension protocol. It inventories all equations, restates each in its own notation, and identifies implementation-relevant details that the PDF leaves implicit: the inverse Mills ratio for numerical stability in truncated normal sampling, a Cholesky precomputation for the Gibbs posterior draw, and an MLE warm-start strategy for the MH chain.

The planner produces four artifacts. No clarification was needed (\textsc{Hold} was not raised).

\begin{artifact}[title={\texttt{comprehension.md}} (excerpt)]
\textbf{Latent variable model:} $y_i^* = \mathbf{x}_i'\boldsymbol{\beta} + \varepsilon_i$, \enspace $\varepsilon_i \sim \mathcal{N}(0,1)$, \enspace $y_i = \mathds{1}(y_i^* > 0)$.\\[4pt]
\textbf{MLE:} Newton-Raphson on $\ell(\boldsymbol{\beta}) = \sum_i \bigl[y_i \log\Phi(\mathbf{x}_i'\boldsymbol{\beta}) + (1-y_i)\log(1-\Phi(\mathbf{x}_i'\boldsymbol{\beta}))\bigr]$. \\
Gradient uses $\phi/\Phi$ (inverse Mills ratio) for numerical stability.\\[4pt]
\textbf{Gibbs (Albert-Chib):} Augment with $y_i^* \mid y_i, \boldsymbol{\beta} \sim \mathcal{TN}(\mathbf{x}_i'\boldsymbol{\beta}, 1)$, truncated to $(0,\infty)$ if $y_i=1$, $(-\infty, 0]$ if $y_i=0$. Then $\boldsymbol{\beta} \mid \mathbf{y}^* \sim \mathcal{N}(\cdot, \cdot)$ via Cholesky.\\[4pt]
\textbf{Self-test verdict:} \textsc{Fully Understood}. All formulas reproduced from memory. Implicit assumption identified: MH proposal scale requires tuning for acceptance $\approx 0.40$.
\end{artifact}

\begin{artifact}[sidebyside, sidebyside align=top, righthand width=0.48\textwidth,
  title={\texttt{spec.md} $\to$ \abuilder{} \hfill \texttt{test-spec.md} $\to$ \atester{}}]
\textbf{Implement} three C++ functions via Rcpp/Armadillo:\\[2pt]
{\footnotesize
\texttt{probit\_mle()}: Newton-Raphson, observed info, warm-start from $\mathbf{0}$\\
\texttt{probit\_gibbs()}: Albert-Chib, 3000 draws, 500 burn-in, Cholesky\\
\texttt{probit\_mh()}: RW-MH, 10000 draws, adaptive proposal
}\\[4pt]
R wrappers: \texttt{probit(method=...)} dispatching to C++.
\tcblower
\textbf{Validate} without seeing implementation:\\[2pt]
{\footnotesize
MLE vs \texttt{glm(link="probit")}: tol $< 10^{-6}$\\
Gibbs posterior mean $\to$ MLE as $n \to \infty$\\
MH acceptance rate $\in [0.25, 0.50]$\\
Edge cases: perfect separation, $p > n/10$, single predictor
}\\[4pt]
Sim validation: bias $\to 0$, coverage $\to 0.95$.
\end{artifact}

\subsection{Parallel Execution}
\label{sec:tutorial-execution}

\aleader{} dispatches three agents in parallel, each receiving only its own specification. \abuilder{} implements three C++ estimation functions using Armadillo linear algebra, with R wrapper functions providing a consistent API. It does not know the ground-truth parameters or the DGP that will be used to evaluate its code. \asimulator{} implements the DGP with known ground-truth $\beta_0 = -1$, $\beta_1 = 0.5$, runs 6,000 total fits ($4 \times 500 \times 3$ methods), and produces comparison tables. It does not know how the estimators are implemented. \atester{} validates whether the implementation recovers ground truth from simulated data: MLE output against R's \texttt{glm()} at $10^{-6}$ tolerance, Bayesian posterior convergence, edge cases. It sees neither the source code nor the DGP design.

After all three pipelines complete, \ascriber{} documents the process and \areviewer{} cross-audits all outputs. Upon passing tests, \ashipper{} commits and pushes to GitHub upon user authorization.

\begin{passsignal}
Pipeline isolation verified. Convergence confirmed: MLE matches \texttt{glm()} at $10^{-8}$. Monte Carlo acceptance criteria satisfied (7/7). Tolerance integrity: no inflation detected. \textbf{Verdict: \textsc{Pass}.}
\end{passsignal}

\subsection{Results}
\label{sec:tutorial-results}

\begin{figure}[!ht]
\centering
\includegraphics[width=0.8\textwidth]{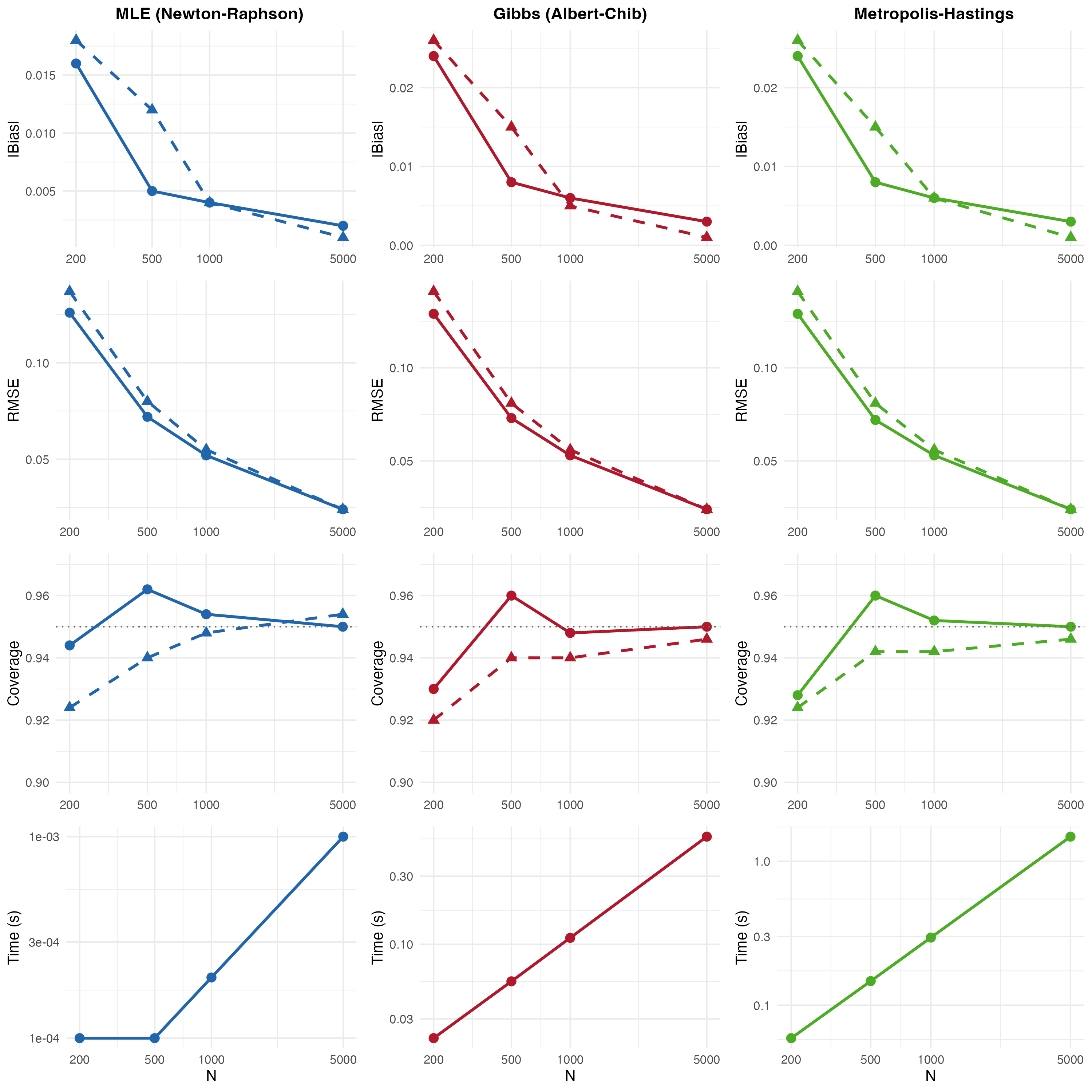}
\caption{Monte Carlo comparison of three probit estimators across different sample sizes $N \in \{200, 500, 1000, 5000\}$ with 500 replications per scenario. Columns: MLE (blue), Gibbs (red), MH (green). Rows: $|\text{Bias}|$, RMSE, 95\% CI coverage, computation time. All three methods exhibit consistency, $\sqrt{N}$-convergence, and nominal coverage---confirming that the C++ implementations match their mathematical specifications.}\label{fig:probit-mc}
\end{figure}

Figure~\ref{fig:probit-mc} presents the Monte Carlo results. All three methods exhibit the theoretically expected behavior: bias converges to zero (consistency), RMSE scales as $1/\sqrt{N}$ ($\sqrt{N}$-convergence), and 95\% CI coverage converges to the nominal level. MLE runs in microseconds; Gibbs and MH require orders of magnitude more time but provide full posterior distributions. MH acceptance rates of 0.396--0.401 are near the $\sim$0.40 theoretical optimum.

The key observation is what the user provided versus what the system produced. The user supplied approximately 50 words of domain guidance and a 4-page PDF. The system produced a complete, installable R package with Rcpp/Armadillo C++ backends, a comprehensive test suite, a Monte Carlo simulation harness, and full documentation. Every substantive decision---which estimation methods, which C++ library, what simulation design---came from the user's prompt and the PDF. The engineering labor---code structure, build system, test infrastructure, simulation harness---was handled by the agents.

\FloatBarrier
\section{Practical Applications}
\label{sec:applications}

The original motivation for developing \textsc{StatsClaw} was practical: the authors maintain several R packages for causal inference and panel data analysis (\texttt{interflex}, \texttt{fect}, \texttt{panelView}), and the engineering cost of keeping these packages tested, documented, and up to date had become the binding constraint on the research. We used \textsc{StatsClaw} to do exactly this work. The three applications below are not constructed demonstrations; they are the actual development tasks that the workflow was built to handle. Each represents a distinct mode of statistical software development---greenfield construction, sustained refactoring, and paper-to-feature---and together they serve as direct testimony to the workflow's efficiency and effectiveness (Table~\ref{tab:applications-comparison}).

\begin{table}[ht]
\centering
\small
\caption{Summary of applications.}\label{tab:applications-comparison}
\resizebox{0.8\textwidth}{!}{%
\begin{tabular}{lccc}
\toprule
\textbf{Metric} & \textbf{\texttt{panelView}} & \textbf{\texttt{interflex} Python} & \textbf{\texttt{fect}} \\
& {\scriptsize (Paper$\to$Feature)} & {\scriptsize (Code Translation)} & {\scriptsize (Refactoring)} \\
\midrule
Duration & Multi-session & 3 rounds & 5 days, ${\sim}$20 runs \\
User input & Prompt + paper & $\sim$150 words & $\sim$10 interactions \\
Test suite & $0 \to$ full suite & $0 \to 34$ & $131 \to 590$ \\
Code scope & Refactored + new & ${\sim}$3{,}500 lines & 100+ files \\
Bugs found & 3 (deprecation) & 6 (2 silent) & 11+ \\
Key challenge & Paper comprehension & Spec. extraction & Dependencies \\
\bottomrule
\end{tabular}%
}
\end{table}

\subsection{Paper to Feature: \texttt{panelView} Network Visualization}
\label{sec:app-panelview}

The \texttt{panelView} R package \citep{mou2024panelview} visualizes panel data through a single-export function with approximately fifty parameters. The task required reading \citet{correia2016feasible}, understanding the bipartite graph representation of panel observation patterns, and adding a \texttt{type = "network"} pathway: constructing the bipartite graph, identifying singletons (degree-1 nodes), drawing convex hulls around connected components, and extending to $k$-partite graphs for multi-way fixed effects.

This application illustrates a distinct mode: \aplanner{} read a methodological paper not to implement the estimator itself, but to extract a visualization concept from the paper's exposition. The bipartite graph in \citet{correia2016feasible} is explanatory apparatus, not a software specification. Translating it into a feature required comprehending the mathematics (incidence matrices, graph connectivity, iterative pruning) and then designing a user-facing API.

\begin{figure}[!ht]
\centering
\begin{minipage}{0.48\textwidth}
\centering
\includegraphics[width=\textwidth]{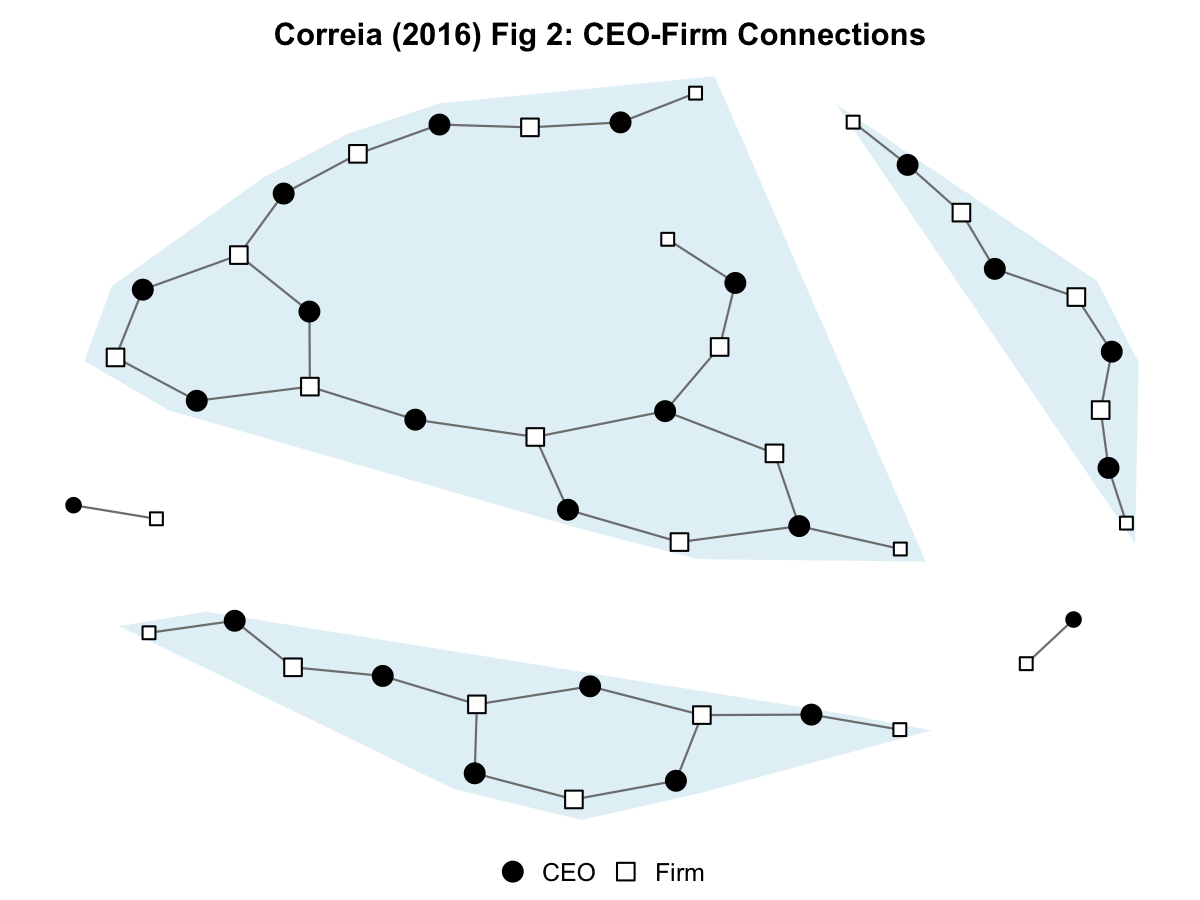}
\end{minipage}\hfill
\begin{minipage}{0.48\textwidth}
\centering
\includegraphics[width=\textwidth]{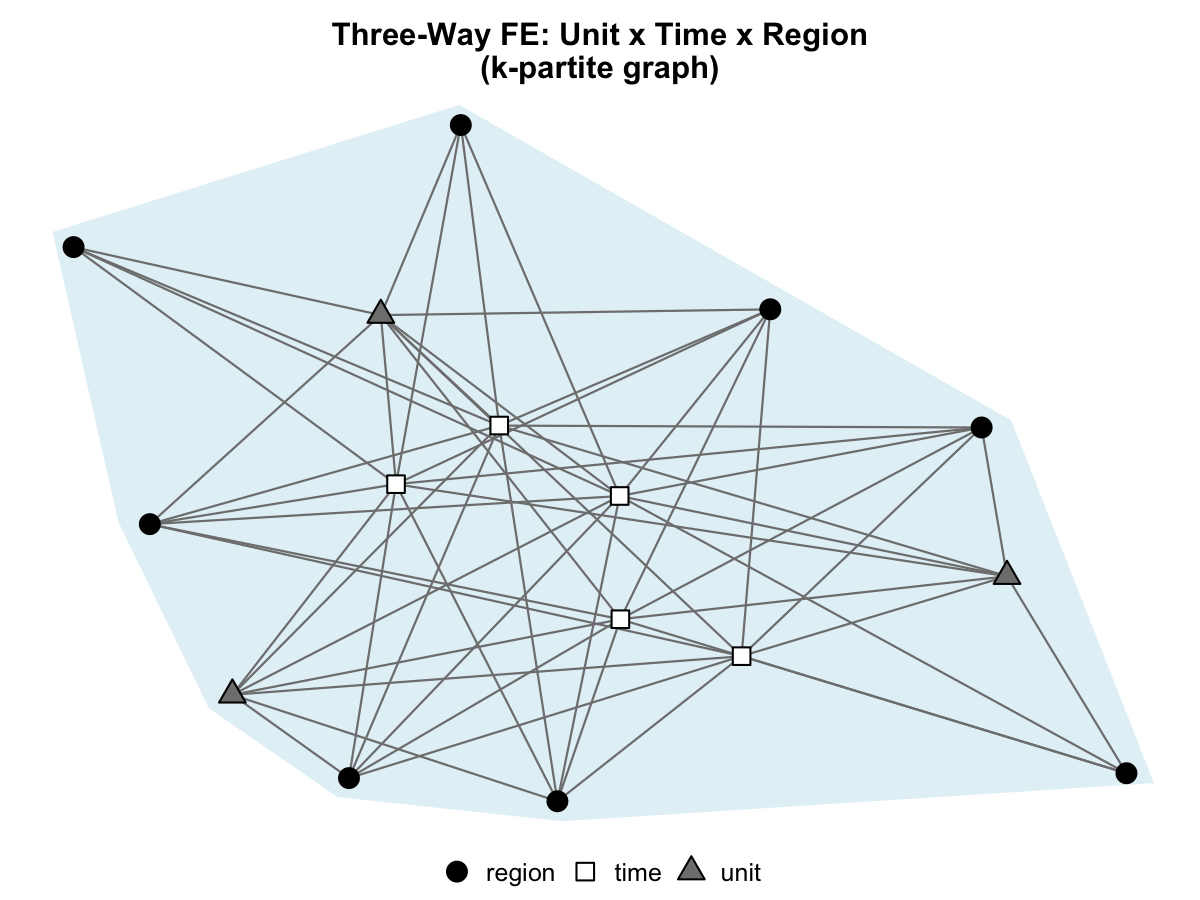}
\end{minipage}
\caption{Left: CEO--Firm bipartite network with 48 nodes, 5 connected components, and 11 singletons, reproducing the diagnostic in \citet{correia2016feasible}. Right: three-way FE (unit $\times$ time $\times$ region) as a $k$-partite graph. Both produced by \texttt{panelview(type = "network")}.}\label{fig:panelview}
\end{figure}

The planner also identified a prerequisite: the existing codebase was a monolithic 37\,KB file. Adding a new dispatch pathway into this monolith would be fragile and difficult to test. The system made a proactive judgment: \textit{refactor first, then add the feature}. The monolith was split into focused modules, a test suite was added, three \texttt{ggplot2} deprecation bugs were fixed, and a 5-chapter Quarto manual replaced the stale vignette---all before the network feature was specified.

Figure~\ref{fig:panelview} shows the results: a CEO--Firm bipartite network reproducing \citet{correia2016feasible}'s key diagnostic, and a $k$-partite extension for three-way fixed effects.

\FloatBarrier

\subsection{Code Translation: \texttt{interflex} for Python}
\label{sec:app-greenfield}

We constructed a complete Python \texttt{interflex} package from the R implementation---zero pre-existing Python code to a validated product with 14 modules, approximately 3,500 lines, 34 tests, and a 10-chapter Quarto tutorial.

No formal mathematical document was provided. \aplanner{} reverse-engineered all contracts from R source code: treatment type polymorphism (auto-detecting discrete vs.\ continuous treatment from the number of unique values), the DML estimation pipeline via DoubleML's influence function and B-spline best linear projection \citep{bach2022doubleml}, and all covariance estimator variants (HC1, homoscedastic, cluster-robust CR1, PCSE).

The user provided approximately 150 words across three rounds: an initial directive to translate the R package, an API refinement (rejecting \texttt{import interflex as ifx} in favor of \texttt{import interflex; interflex(data, ...)}), and a quality assurance request.

\paragraph*{The six-bug audit.} The most significant outcome was not the initial construction but what happened after the user's third message triggered a comprehensive audit. \areviewer{}'s cross-pipeline analysis uncovered six bugs in code that already passed all 34 tests: (1)~HC1 sandwich operator precedence, where NumPy's \texttt{*} binds before \texttt{@}, producing a semantically incorrect covariance computation; (2)~discrete bootstrap silent fallthrough, where \texttt{elif vartype == "bootstrap": pass} silently returned delta method standard errors instead of bootstrap SEs---\textbf{silently wrong statistical results}; (3)~binning weight slicing, where DataFrame operations on a NumPy array caused a crash on any weighted binning regression; (4)~a dead import that caused \texttt{ImportError} on any kernel call; (5)~a plotting distribution overlay that used 50 evaluation grid points instead of raw moderator values; and (6)~dead prediction code that claimed B-spline re-projection but performed generic interpolation.

\begin{figure}[ht]
\centering
\begin{minipage}{0.48\textwidth}
\centering
\includegraphics[width=\textwidth]{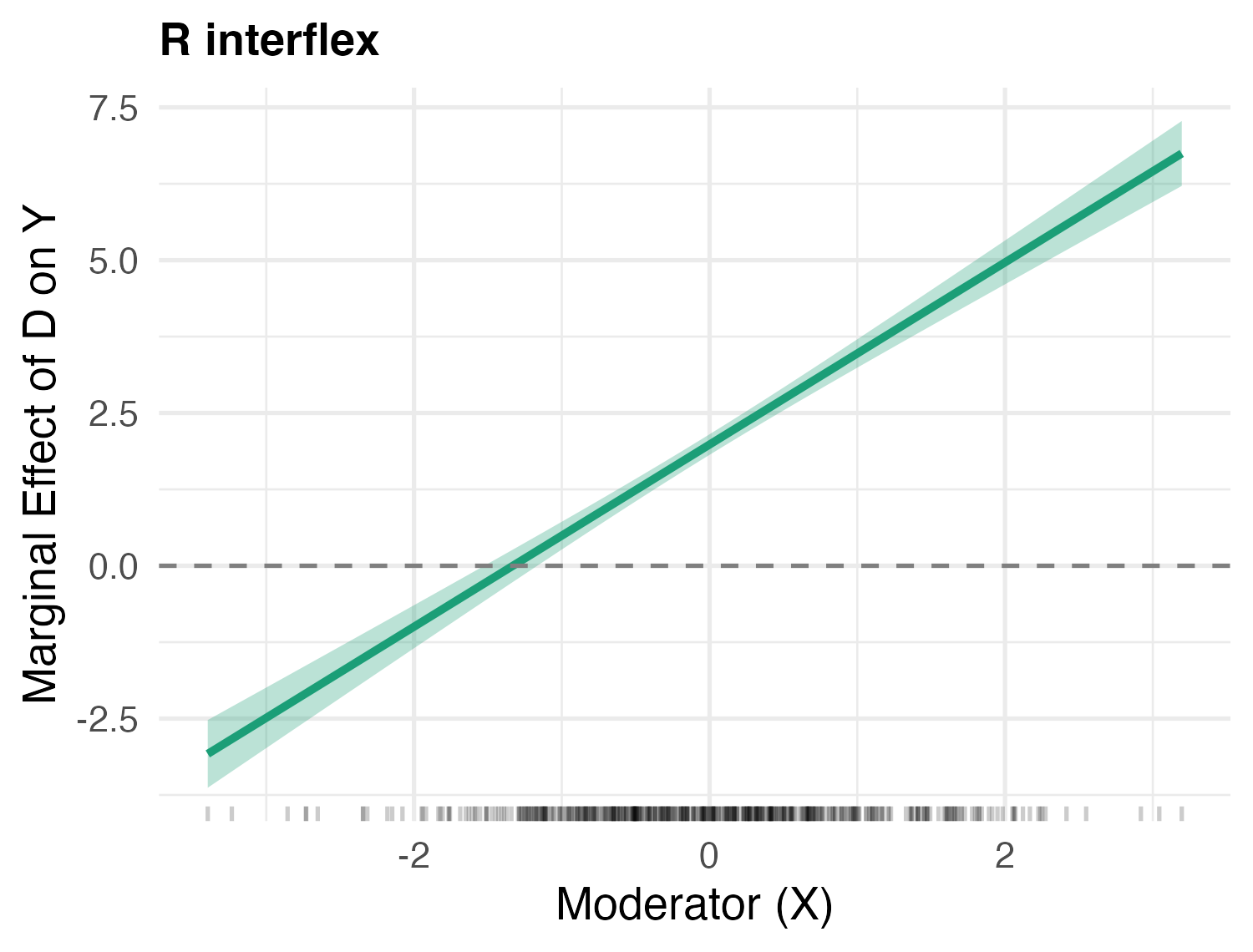}
\end{minipage}\hfill
\begin{minipage}{0.48\textwidth}
\centering
\includegraphics[width=\textwidth]{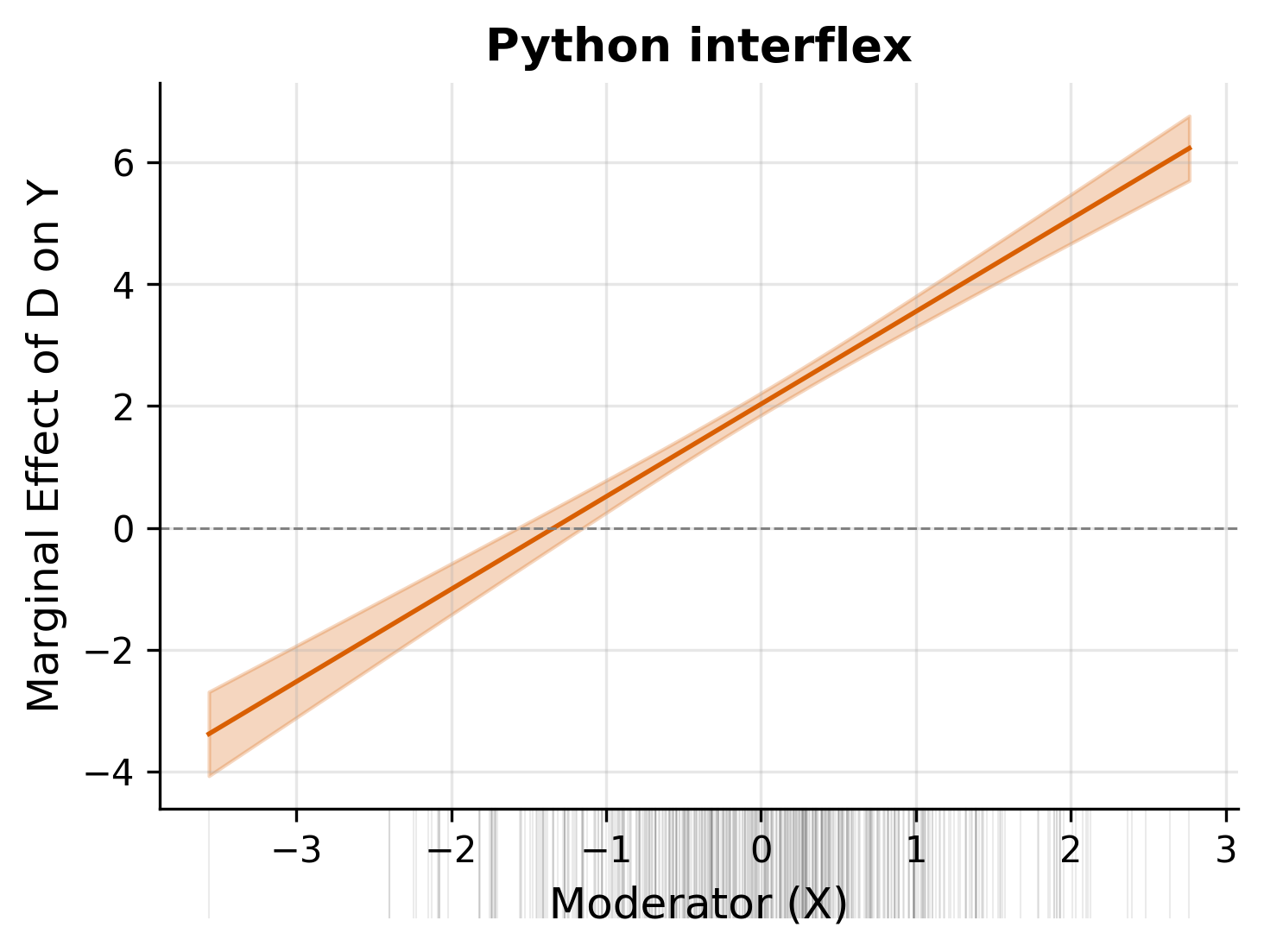}
\end{minipage}
\caption{Conditional marginal effect estimates from R \texttt{interflex} (left) and Python \texttt{interflex} (right) on the same DGP. Both recover the true conditional marginal effect $\partial E[Y|D,X]/\partial D = 2 + 1.5X$ with matching point estimates and confidence intervals.}\label{fig:interflex-comparison}
\end{figure}

Figure~\ref{fig:interflex-comparison} confirms visual and numerical equivalence between the R and Python implementations.

Bugs~(2) and~(3) would have produced silently incorrect inferential results under specific parameter combinations. Their detection by the reviewer's cross-pipeline audit---not by the test suite---is the strongest evidence in this paper for the value of separated verification. A generate-then-test workflow, where the same model writes both the code and the tests, would likely have embedded the same misunderstanding in both.

\FloatBarrier

\subsection{Sustained Development: \texttt{fect}}
\label{sec:app-fect}

The \texttt{fect} R package \citep{liu2024practical} implements a group of counterfactual estimators, including interactive fixed effects, matrix completion, and complex fixed effects, for causal panel analysis. A five-day refactoring campaign (approximately twenty workflow runs) addressed six interdependent workstreams: structural refactoring, cross-validation unification, convergence conditioning in C++ solvers, visualization overhaul, a 12-chapter user manual, and bug resolution. The user provided approximately ten substantive interactions covering methodology, DGP design, and documentation structure.

This is the most demanding application because changes compound: modifying a cross-validation scoring function affects parameter interfaces, bootstrap routines, plotting code, and documentation. Missing a single dependency produces bugs that may be silent for months.

\paragraph*{Convergence conditioning.} The IFE model decomposes the outcome matrix as $Y = \mu \mathbf{1}\mathbf{1}' + \alpha \mathbf{1}' + \mathbf{1}\xi' + F\Lambda' + \varepsilon$. The EM solver tracked convergence via $\|Y^{(t)} - Y^{(t-1)}\|_F / \|Y^{(t-1)}\|_F$. When $\mu$ is large (e.g., GDP data, $\mu \sim 10^4$), this global criterion is dominated by the grand mean: at $\texttt{tol} = 10^{-3}$, the solver converged with 9.4\% relative error in $F\Lambda'$. \aplanner{} specified a two-part fix: R-level centering (subtracting $\hat{\mu}$ before the C++ call) and C++ component-wise convergence monitoring. The results were dramatic:

\begin{center}
\small
\begin{tabular}{lccc}
\toprule
\textbf{Component} & \textbf{Old error} & \textbf{New error} & \textbf{Improvement} \\
\midrule
$\hat{\alpha}$ (unit FE) & $2.03 \times 10^{-1}$ & $3.72 \times 10^{-3}$ & 55$\times$ \\
$FL'$ (factors) & $2.99$ & $6.89 \times 10^{-2}$ & 43$\times$ \\
$\hat{\sigma}^2$ & $5.07 \times 10^{-3}$ & $2.25 \times 10^{-6}$ & 2{,}249$\times$ \\
\bottomrule
\end{tabular}
\end{center}

\begin{figure}[ht]
\centering
\includegraphics[width=0.7\textwidth]{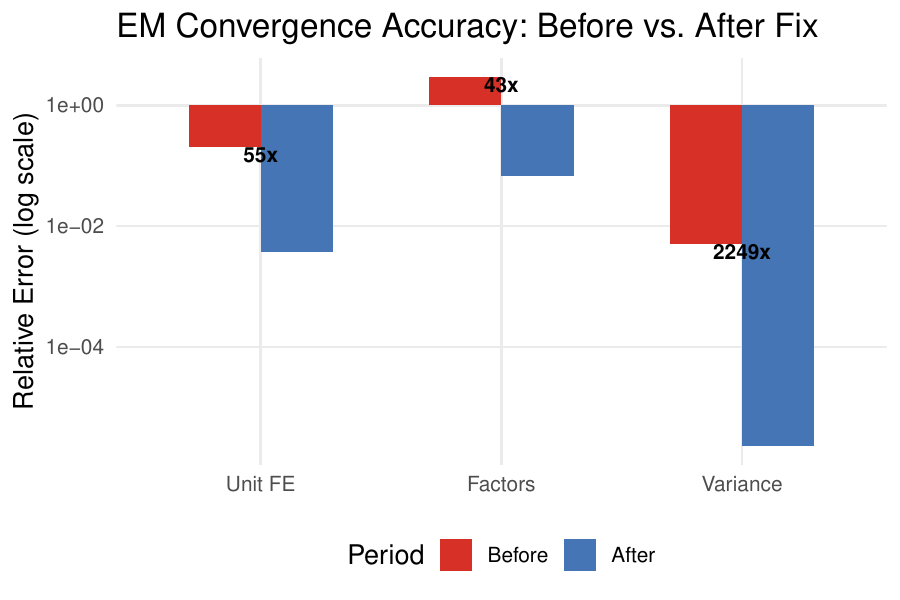}
\caption{Component-wise relative error before and after the convergence fix ($\texttt{tol}=10^{-3}$). The old global criterion allowed 9.4\% error in the factor component because the grand mean dominated the denominator. The new criterion ensures each component converges to its own scale.}\label{fig:fect-convergence}
\end{figure}

This is not just an engineering fix. Premature convergence in the EM solver propagates directly into ATT estimates, potentially invalidating published analyses that use \texttt{fect} on data with large level effects. Figure~\ref{fig:fect-convergence} quantifies the improvement.

\paragraph*{Adversarial testing as discovery.} The most scientifically significant finding emerged on Day~4 of the campaign. The builder implemented a data-generating process for the complex fixed effects chapter on a balanced panel; the tester's placebo test showed no improvement from additional fixed effects. The second-dispatch builder identified the root cause: on a balanced panel, the two-way demeaning operator $M_D = I - D(D'D)^{-1}D'$ absorbs group structure entirely. If group membership is collinear with the unit dimension, the residual group mean is algebraically zero for all groups, making the extra FE a no-op regardless of confounding strength.

This mathematical insight---that balanced-panel geometry renders certain fixed-effect augmentations algebraically degenerate---was not anticipated by the planner or the builder. It was surfaced by the tester's behavioral observation:

\begin{blocksignal}
Placebo test failure: adding group FE to balanced panel produces zero improvement in RMSE. Expected $>$5\% reduction under DGP with group confounding. Suspecting algebraic degeneracy---requesting builder investigation.
\end{blocksignal}

\noindent The separated verification architecture did not merely catch a bug; it discovered a property of the estimator that the developers had not considered. This is the strongest qualitative evidence that adversarial verification can serve as a discovery mechanism, not just a quality assurance tool.

\begin{figure}[!ht]
\centering
\includegraphics[width=0.7\textwidth]{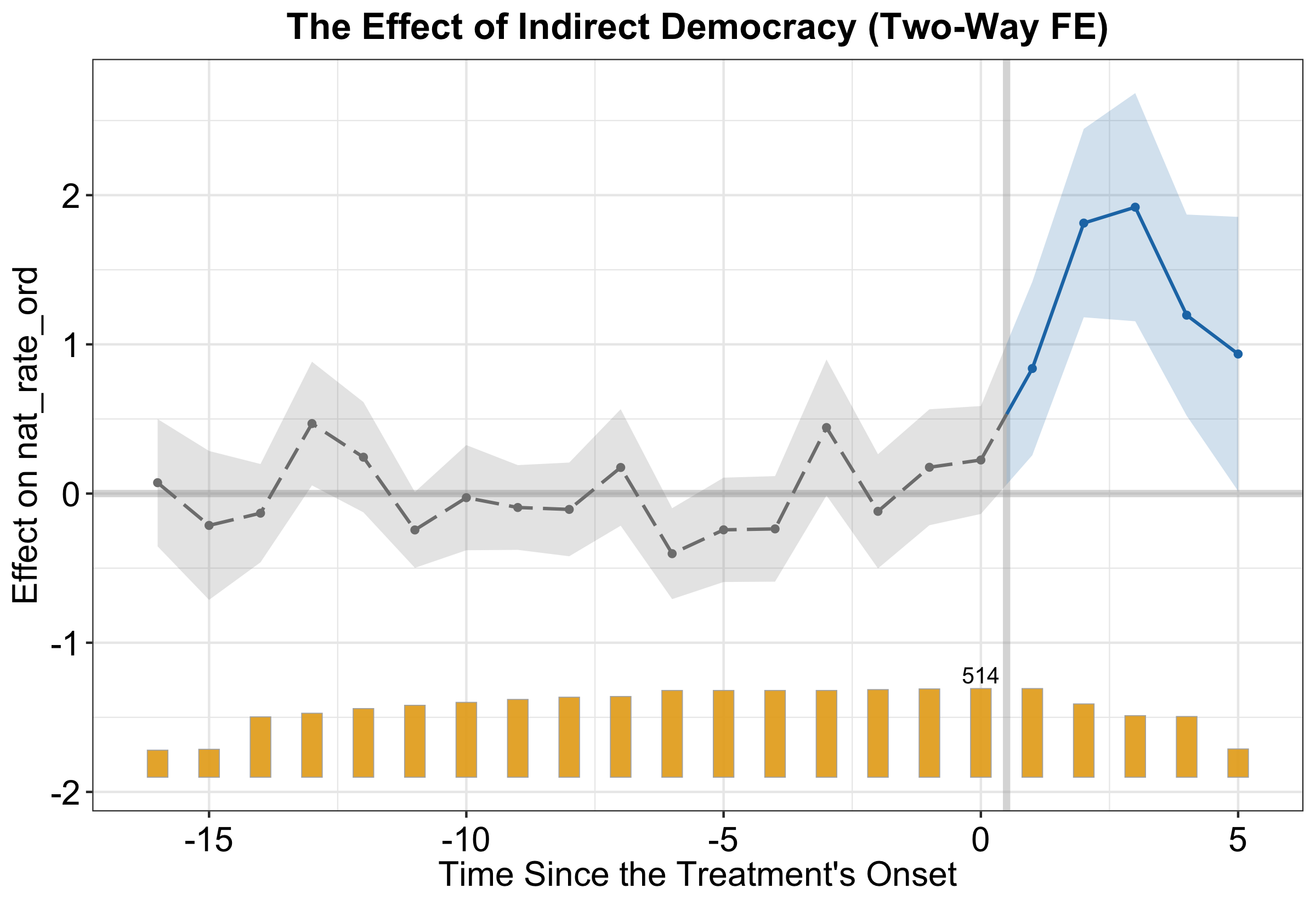}
\caption{The effect of indirect democracy on naturalization rates \citep{hainmueller2019indirect}, estimated by two-way FE via \texttt{fect} (500 bootstrap replications). Grey ribbon: pre-treatment ATT centered at zero across 15 pre-treatment periods (parallel-trends validation). Blue ribbon: post-treatment effect rising to $+1.8$ at $t+2$. Golden bars: number of treated units at each relative period.}\label{fig:fect-democracy}
\end{figure}

\paragraph*{Feature improvement.} Figure~\ref{fig:fect-democracy} illustrates one outcome of the refactoring: the visualization of treatment effects from the \citet{hainmueller2019indirect} study of Swiss municipalities' switch from direct to indirect democracy. Beyond the statistical results, the figure demonstrates the workflow's ability to handle delicate visual details through iterative builder--tester cycles with a human in the loop. The smooth color transition from grey (pre-treatment) to blue (post-treatment) at $t = 0$---a rendering detail that would typically require hours of manual coding to get right---was refined through successive \textsc{Block} and fix rounds until the visualization met the user's specifications.

\paragraph*{Quantitative summary.} Over five days: 33+ commits, tests grew from 131 to 590 (100\% pass rate), 11+ \textsc{Block} signals raised and resolved, 12-chapter Quarto manual produced, and zero known bugs at completion.

\FloatBarrier

\section{Discussion}
\label{sec:discussion}

The demonstration and applications tested the workflow under progressively harder conditions: from a textbook estimation problem with known ground truth (Section~\ref{sec:tutorial}), through three practical applications (Section~\ref{sec:applications}). We now assess what these experiences reveal about the workflow's strengths, limitations, and broader implications.

\subsection{What Worked}

Three patterns emerged consistently across applications. First, separated verification catches what testing misses. The six-bug audit in the Python \texttt{interflex} package is the clearest example: two bugs that would have produced silently incorrect statistical results survived a 34-test suite but were caught by the reviewer's cross-pipeline analysis (Figure~\ref{fig:interflex-comparison}). The key insight is that tests written by the same process that wrote the code share the same blind spots. Independent specification breaks this correlation.

Second, adversarial testing can serve as a discovery mechanism. The balanced-panel degeneracy in \texttt{fect} was not a bug but a mathematical property of the estimator that the developers had not considered. It was surfaced by the tester's behavioral observation through a \textsc{Block} signal---not by deliberate investigation (Figure~\ref{fig:fect-convergence}). When two independently derived specifications disagree, the disagreement reveals not only bugs but also gaps in understanding.

Third, the comprehension protocol routes human attention efficiently. The \texttt{panelView} network visualization (Figure~\ref{fig:panelview}) illustrates this most clearly: the planner read a methodological paper, extracted a visualization concept, and proactively refactored a monolithic codebase---all from a single user prompt. The \textsc{Hold} mechanism channels user engagement toward domain decisions---which covariance estimator? which DGP design? which tolerance is scientifically meaningful?---while delegating engineering mechanics to the agents. The user's cognitive load scales with the domain complexity of the task, not with the engineering complexity. More broadly, our experience suggests that the quality of the workflow's output scales with the depth of human involvement: the more discussion between the user and the agents during planning, the better the specifications; the more the user participates in designing test criteria, the more effective the execution. The workflow substantially reduces the cost of package maintenance and new feature development, but it should not be mistaken for a tool that automatically produces perfect software from a single prompt.

\subsection{Limitations}

The quality of output is ultimately bounded by the underlying language model's capability. When source material is dense and highly abstract---for example, a manuscript that assumes familiarity with advanced functional analysis or measure-theoretic probability---the planner's comprehension may be shallow, producing specifications that miss subtle requirements. In such cases, performance degrades and the user must compensate with more detailed guidance. Even for material within the model's reach, methods requiring extremely deep domain-specific reasoning may exceed current capabilities, triggering repeated \textsc{Hold} cycles. These cycles are a feature---they channel expert input to where it matters---but they impose a practical limit on autonomy.

More fundamentally, the adversarial architecture provides high confidence but not formal correctness guarantees. It detects bugs empirically through independent behavioral testing, not through mathematical proof. Systematically silent misunderstandings---where the model is confidently wrong across all pipelines---remain a residual risk. The comprehension protocol mitigates this by creating a single auditable checkpoint, but it requires a domain expert who can assess whether the system's mathematical understanding is correct. Users who lack this expertise will not benefit from the protocol. The system amplifies expertise; it does not replace it.

Finally, multi-pipeline execution with independent builder, tester, and simulator agents incurs higher computational cost than single-pass code generation. This tradeoff favors correctness over speed---appropriate for statistical software where a silent bug can invalidate published analyses, but worth acknowledging as a constraint on routine use.


\subsection{Implications}

For methodological researchers, the engineering bottleneck documented by \citet{ram2019community} need no longer be binding. A small team---even a single researcher---can produce tested, documented, distributable software at a pace that previously required dedicated engineering resources. This does not change the science; it changes the logistics of how science reaches practitioners.

For the broader statistical computing community, the workflow's process recording creates audit trails by default \citep{gentleman2007statistical, stodden2016enhancing}. Every decision, every test result, every comprehension check is documented in a machine-readable workspace repository. This is not a substitute for formal reproducibility standards \citep{peng2011reproducible}, but it provides a layer of transparency that is absent from most software development workflows.

For software citation \citep{smith2016software}, the workflow lowers the barrier to producing citable, version-controlled packages---making it easier for methodological contributions to receive credit through their software implementations.

\section{Getting Started}
\label{sec:getting-started}

\textsc{StatsClaw} requires two components: Claude Code (Anthropic's command-line agent) and a GitHub-hosted target repository with push access. No separate installation is needed---the framework is loaded as Claude Code's system configuration.

\paragraph*{Setup.} Three steps:

\begin{terminal}
\textcolor{termaccent}{\$} git clone https://github.com/statsclaw/statsclaw.git
\textcolor{termaccent}{\$} cd statsclaw
\textcolor{termaccent}{\$} claude                \textcolor{gray}{opens Claude Code with StatsClaw loaded}
\end{terminal}

\noindent The agent reads the configuration automatically. Optionally, create a workspace repository (a separate GitHub repo) to store process records and handoff notes across sessions.

\paragraph*{Usage.} Provide a natural-language directive specifying the target project and desired outcome. \aleader{} resolves the repository, detects the language (R, Python, Stata, Go, Rust, C/C++, TypeScript), selects the appropriate workflow, and proceeds autonomously---raising \textsc{Hold} signals when user input is needed.

\paragraph*{Example prompts.}

\begin{center}
\small
\begin{tabular}{ll}
\toprule
\textbf{Prompt} & \textbf{Workflow} \\
\midrule
\textit{``Fix the failing tests in fect''} & Code change \\
\textit{``Build a Python package from this R package''} & Code + ship \\
\textit{``Update the README and vignette''} & Docs only \\
\textit{``Run a Monte Carlo comparing these estimators''} & Simulation study \\
\textit{``Read this paper and add the feature to panelView''} & Code + ship \\
\bottomrule
\end{tabular}
\end{center}

\paragraph*{Effective patterns.} Provide domain guidance, not engineering instructions. ``Use HC1 with cluster-robust standard errors'' is more useful than ``Create a function called \texttt{compute\_se()}.'' Review the \texttt{comprehension.md} artifact before approving code generation---a five-minute review prevents hours of rework. Respond to \textsc{Hold} signals with precision: ``Use \texttt{lightgbm}'s defaults'' is better than ``Do whatever seems reasonable.''

Full architecture details, the ten workflow types, and language-specific profiles are described in Appendix~\ref{sec:appendix}.

\paragraph*{Contributing.} \textsc{StatsClaw} is an open-source project, and we welcome contributions from the community. After using the workflow on your own packages, you can run the built-in \texttt{/contribute} command, which summarizes the lessons learned during your session---what worked, what required manual intervention, and what domain-specific patterns emerged---into a structured report. With your permission, these reports (\texttt{.md} files) can be submitted to the GitHub repository, where they may be absorbed into the framework's configuration to improve future performance for similar tasks. We invite researchers to visit \url{https://statsclaw.ai/} or join us at \url{https://github.com/statsclaw/statsclaw} and contribute in whatever way they can.


\vspace{3em}
\singlespacing
\bibliographystyle{plainnat}
\bibliography{refs}

@article{mccullough1999numerical,
  author  = {McCullough, B. D. and Vinod, H. D.},
  title   = {The Numerical Reliability of Econometric Software},
  journal = {Journal of Economic Literature},
  volume  = {37},
  number  = {2},
  pages   = {633--665},
  year    = {1999},
  doi     = {10.1257/jel.37.2.633}
}

@article{ram2019community,
  author  = {Ram, Karthik and Boettiger, Carl and Chamberlain, Scott and Ross, Noam and Goldberg, Maelle and Bartomeus, Ignasi},
  title   = {A Community of Practice Around Peer Review for Long-Term Research Software Sustainability},
  journal = {Computing in Science \& Engineering},
  volume  = {21},
  number  = {1},
  pages   = {59--65},
  year    = {2019},
  doi     = {10.1109/MCSE.2018.2882753}
}

@article{peng2011reproducible,
  author  = {Peng, Roger D.},
  title   = {Reproducible Research in Computational Science},
  journal = {Science},
  volume  = {334},
  number  = {6060},
  pages   = {1226--1227},
  year    = {2011},
  doi     = {10.1126/science.1213847}
}

@article{chen2021evaluating,
  author  = {Chen, Mark and Tworek, Jerry and Jun, Heewoo and Yuan, Qiming and Pinto, Henrique Ponde de Oliveira and Kaplan, Jared and Edwards, Harri and Burda, Yuri and Joseph, Nicholas and Brockman, Greg and others},
  title   = {Evaluating Large Language Models Trained on Code},
  journal = {arXiv preprint arXiv:2107.03374},
  year    = {2021}
}

@inproceedings{jimenez2024swebench,
  author    = {Jimenez, Carlos E. and Yang, John and Wettig, Alexander and Yao, Shunyu and Pei, Kexin and Press, Ofir and Narasimhan, Karthik},
  title     = {{SWE-bench}: Can Language Models Resolve Real-World {GitHub} Issues?},
  booktitle = {International Conference on Learning Representations},
  year      = {2024}
}

@inproceedings{yang2024sweagent,
  author    = {Yang, John and Jimenez, Carlos E. and Wettig, Alexander and Lieret, Kilian and Yao, Shunyu and Narasimhan, Karthik and Press, Ofir},
  title     = {{SWE-agent}: Agent-Computer Interfaces Enable Automated Software Engineering},
  booktitle = {Advances in Neural Information Processing Systems},
  year      = {2024}
}

@article{gentleman2007statistical,
  author  = {Gentleman, Robert and Temple Lang, Duncan},
  title   = {Statistical Analyses and Reproducible Research},
  journal = {Journal of Computational and Graphical Statistics},
  volume  = {16},
  number  = {1},
  pages   = {1--23},
  year    = {2007},
  doi     = {10.1198/106186007X178663}
}

@article{stodden2016enhancing,
  author  = {Stodden, Victoria and McNutt, Marcia and Bailey, David H. and Deelman, Ewa and Gil, Yolanda and Hanson, Brooks and Heroux, Michael A. and Ioannidis, John P. A. and Taufer, Michela},
  title   = {Enhancing Reproducibility for Computational Methods},
  journal = {Science},
  volume  = {354},
  number  = {6317},
  pages   = {1240--1241},
  year    = {2016},
  doi     = {10.1126/science.aah6168}
}

@article{smith2016software,
  author  = {Smith, Arfon M. and Katz, Daniel S. and Niemeyer, Kyle E.},
  title   = {Software Citation Principles},
  journal = {PeerJ Computer Science},
  volume  = {2},
  pages   = {e86},
  year    = {2016},
  doi     = {10.7717/peerj-cs.86}
}

@article{bach2022doubleml,
  title     = {{DoubleML}---An Object-Oriented Implementation of Double Machine Learning in {Python}},
  author    = {Bach, Philipp and Chernozhukov, Victor and Kurz, Malte S. and Spindler, Martin},
  journal   = {Journal of Machine Learning Research},
  volume    = {25},
  number    = {96},
  pages     = {1--8},
  year      = {2024}
}

@article{liu2024practical,
  title     = {Practical Causal Inference with Panel Data},
  author    = {Liu, Licheng and Wang, Ye and Xu, Yiqing},
  journal   = {Journal of the American Statistical Association},
  year      = {2024},
  note      = {Forthcoming},
  doi       = {10.1080/01621459.2024.2395588}
}

@article{mou2024panelview,
  title     = {{panelView}: Visualizing Panel Data},
  author    = {Mou, Hongyu and Liu, Licheng and Xu, Yiqing},
  journal   = {Journal of Statistical Software},
  volume    = {107},
  number    = {7},
  pages     = {1--20},
  year      = {2024},
  doi       = {10.18637/jss.v107.i07}
}

@article{correia2016feasible,
  title     = {A Feasible Estimator for Linear Models with Multi-Way Fixed Effects},
  author    = {Correia, Sergio},
  year      = {2016},
  note      = {Working paper, Duke University}
}

@article{hainmueller2019indirect,
  title     = {Does Direct Democracy Hurt Immigrant Minorities? {E}vidence from Naturalization Decisions in {Switzerland}},
  author    = {Hainmueller, Jens and Hangartner, Dominik},
  journal   = {American Journal of Political Science},
  volume    = {63},
  number    = {3},
  pages     = {530--551},
  year      = {2019},
  doi       = {10.1111/ajps.12433}
}

@article{albert1993bayesian,
  title     = {Bayesian Analysis of Binary and Polychotomous Response Data},
  author    = {Albert, James H. and Chib, Siddhartha},
  journal   = {Journal of the American Statistical Association},
  volume    = {88},
  number    = {422},
  pages     = {669--679},
  year      = {1993},
  doi       = {10.1080/01621459.1993.10476321}
}


\clearpage
\appendix

\onehalfspacing
\setcounter{page}{1}
\setcounter{table}{0}
\setcounter{figure}{0}
\setcounter{equation}{0}
\setcounter{footnote}{0}

\renewcommand{\theassumption}{A\arabic{assumption}}
\renewcommand\thetable{A\arabic{table}}
\renewcommand\thefigure{A\arabic{figure}}
\renewcommand{\thepage}{A-\arabic{page}}
\renewcommand{\theequation}{A\arabic{equation}}
\renewcommand{\thefootnote}{A\arabic{footnote}}

\section{\textsc{StatsClaw} Architecture Reference}
\label{sec:appendix}

This appendix provides the complete technical specification of the \textsc{StatsClaw} framework. Sections~\ref{sec:approach} and \ref{sec:tutorial} of the main text describe the principles and demonstrate the workflow; this appendix documents the implementation in full.

\subsection{Agent Specifications}
\label{sec:appendix-agents}

\textsc{StatsClaw} comprises agents dispatched within a single Claude Code session via the built-in \texttt{Agent} tool. Table~\ref{tab:agents} summarizes each agent's role, pipeline assignment, and isolation constraints.

\begin{table}[ht]
\centering
\small
\caption{Agent specifications.}\label{tab:agents}
\begin{tabular}{llll}
\toprule
\textbf{Agent} & \textbf{Role} & \textbf{Pipeline} & \textbf{Isolation} \\
\midrule
Leader    & Orchestrator           & Control      & N/A (dispatches only) \\
Planner   & Requirements analyst   & Bridge       & Sees all sources \\
Builder   & Code implementer       & Code         & Information + worktree \\
Tester    & Independent validator  & Test         & Information + worktree \\
Simulator & Monte Carlo evaluator  & Simulation   & Information + worktree \\
Scriber   & Documentation          & Recording    & Worktree \\
Reviewer  & Quality gate           & Convergence  & Sees all pipelines \\
Shipper   & Distribution           & Ship         & Sees all artifacts \\
\bottomrule
\end{tabular}
\end{table}

\bigskip\noindent\aleader{} is the main orchestrating agent. It resolves the target repository, detects the language profile, selects the workflow type, creates the run directory, and dispatches all other agents in sequence. The leader \textit{never} performs specialist work directly---it does not edit source files, run tests, write documentation, or interact with git. If the leader detects that a task falls within another agent's responsibility, it dispatches that agent rather than acting itself.

\bigskip\noindent\aplanner{} is the sole bridge agent: the only agent with visibility into all source material. It ingests all source material---mathematical derivations, existing codebases, pseudocode, algorithm descriptions---and executes the deep comprehension protocol (Section~\ref{sec:approach-comprehension}); and produces independent specification artifacts for each downstream pipeline. The planner's output is the foundation of the entire workflow---errors here propagate everywhere. This is why the comprehension protocol is a hard gate: the planner must demonstrate understanding before producing specifications.

\bigskip\noindent\abuilder{} receives only \texttt{spec.md} and implements the methodology as source code with internal unit tests. It operates in an isolated git worktree to prevent filesystem conflicts with other agents. The builder never sees \texttt{test-spec.md} or \texttt{sim-spec.md}. Its deliverable is \texttt{implementation.md}, documenting files changed, design choices, and any unit tests written.

\bigskip\noindent\atester{} receives only \texttt{test-spec.md} and constructs an independent validation suite from behavioral contracts, numerical tolerances, edge cases, and property-based invariants. It never sees \texttt{spec.md} or the builder's source code structure. Its deliverable is \texttt{audit.md}, containing a per-test result table, before/after comparison tables, and full command output. The tester is the only agent that can raise a \textsc{Block} signal (Section~\ref{sec:appendix-signals}).

\bigskip\noindent\asimulator{} is activated for the full workflow. The simulator receives only \texttt{sim-spec.md} and implements data-generating processes and finite-sample performance evaluation. It treats the estimator as a black-box interface---calling the implementation without knowledge of its internal structure. Its deliverable is \texttt{simulation.md}, containing DGP implementation details, harness design, smoke-run results, and acceptance criteria assessment.

\bigskip\noindent\ascriber{} produces three mandatory artifacts after the execution pipelines complete. First, \texttt{Architecture.md} contains Mermaid diagrams (module structure, function call graph, data flow) and reference tables, written to both the target repository root and the run directory. Second, \texttt{log-entry.md} contains the complete process record: what changed, implementation notes, validation results, problems encountered, review summary, design decisions, and handoff notes. Third \texttt{docs.md} summarizes documentation changes. The scriber operates in an isolated worktree.

\bigskip\noindent\areviewer{} The reviewer is the second agent with full cross-pipeline visibility. It reads all specification documents, all pipeline outputs, and the scriber's documentation. It performs seven checks: (1) comprehension verification, (2) pipeline isolation (verifying no agent accessed another's specification), (3) cross-comparison of specifications, (4) convergence of pipeline outputs, (5) test coverage of changed code paths, (6) tolerance integrity (ensuring the tester used exact tolerances from \texttt{test-spec.md}, with no inflation), and (7) validation evidence completeness. The reviewer issues one of three verdicts: \textsc{Pass} (ship safe), \textsc{Pass with Note} (ship with documented concern), or \textsc{Stop} (block and route to responsible agent).

\bigskip\noindent\ashipper{} handles all git and GitHub operations. It verifies the reviewer's \textsc{Pass} verdict (hard gate), confirms the target repository remote, stages code changes and user-facing documentation, commits, and pushes. It then synchronizes the workspace repository: copying \texttt{log-entry.md} to the runs archive, updating \texttt{CHANGELOG.md} and \texttt{HANDOFF.md}, and pushing. The shipper acts only upon explicit user authorization.

\subsection{State Machine}
\label{sec:appendix-state}

\begin{figure}[ht]
\centering
\begin{tikzpicture}[
    >=Stealth,
    state/.style={
        rectangle, rounded corners=3pt, minimum width=3.4cm, minimum height=0.7cm,
        font=\sffamily\scriptsize\bfseries, fill=stateblue, draw=controlblue!40, text=controlblue
    },
    signal/.style={
        diamond, minimum width=1.6cm, minimum height=1.2cm,
        font=\sffamily\scriptsize\bfseries, inner sep=1pt
    },
    flowline/.style={->, thick, color=#1},
    dashedflow/.style={->, dashed, thick, color=#1},
    note/.style={font=\sffamily\scriptsize, text=shipgray, align=center},
]

\node[state] (s1) at (0, 8) {CREDENTIALS\_VERIFIED};
\node[state] (s2) at (0, 7) {NEW};
\node[state] (s3) at (0, 6) {PLANNED};
\node[state] (s4) at (0, 5) {SPEC\_READY};
\node[state] (s5) at (0, 4) {PIPELINES\_COMPLETE};
\node[state] (s6) at (0, 3) {DOCUMENTED};
\node[state] (s7) at (0, 2) {REVIEW\_PASSED};
\node[state] (s8) at (0, 1) {READY\_TO\_SHIP};
\node[state, fill=testgreen!20, draw=testgreen!40, text=testgreen!80!black] (s9) at (0, 0) {DONE};

\draw[flowline=controlblue!60] (s1) -- (s2);
\draw[flowline=controlblue!60] (s2) -- (s3);
\draw[flowline=controlblue!60] (s3) -- (s4);
\draw[flowline=controlblue!60] (s4) -- (s5);
\draw[flowline=controlblue!60] (s5) -- (s6);
\draw[flowline=controlblue!60] (s6) -- (s7);
\draw[flowline=controlblue!60] (s7) -- (s8);
\draw[flowline=controlblue!60] (s8) -- (s9);

\node[signal, fill=signalamber, draw=recordamber, text=recordamber!80!black] (hold) at (4, 5.5) {HOLD};
\node[signal, fill=signalred, draw=convergered, text=convergered!80!black] (block) at (4, 3.5) {BLOCK};
\node[signal, fill=signalred!80, draw=convergered!80, text=convergered!80!black] (stop) at (4, 1.5) {STOP};

\draw[dashedflow=recordamber] (s3.east) -- ++(0.5,0) |- (hold.west);
\draw[dashedflow=convergered] (s5.east) -- ++(0.3,0) |- (block.west);
\draw[dashedflow=convergered!80] (s7.east) -- ++(0.5,0) |- (stop.west);

\node[note, right=0.3cm of hold] {Ask user};
\node[note, right=0.3cm of block] {Respawn\\builder};
\node[note, right=0.3cm of stop] {Respawn\\per routing};

\end{tikzpicture}
\caption{State machine with interrupt signals. Three signals can interrupt progression: \textsc{Hold} (request user input), \textsc{Block} (validation failure, respawn builder), and \textsc{Stop} (quality gate failure, respawn per reviewer routing). Each signal permits up to three retry cycles before escalating to the user.}\label{fig:statemachine}
\end{figure}
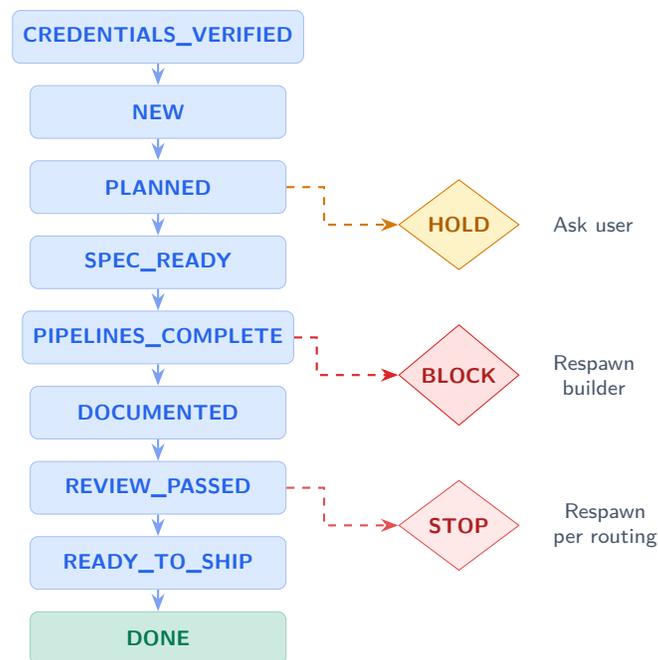

Workflow progression is governed by a state machine with mandatory preconditions enforced as hard gates at each transition (Figure~\ref{fig:statemachine}). Each transition requires specific artifacts to exist. \textsc{Spec\_Ready} requires \texttt{comprehension.md}, \texttt{spec.md}, and \texttt{test-spec.md} from the planner. \textsc{Pipelines\_Complete} requires \texttt{implementation.md} from the builder and \texttt{audit.md} from the tester (and \texttt{simulation.md} from the simulator, if active). \textsc{Documented} requires \texttt{Architecture.md} and \texttt{log-entry.md} from the scriber. \textsc{Review\_Passed} requires a verdict in \texttt{review.md} from the reviewer.

No state can be skipped. The leader checks preconditions before advancing and will not dispatch downstream agents until upstream artifacts are verified.

\subsection{Interrupt Signals}
\label{sec:appendix-signals}

Three interrupt signals govern human--AI interaction and inter-agent error handling:

\begin{table}[ht]
\centering
\small
\caption{Signal handling.}\label{tab:signals}
\begin{tabular}{llll}
\toprule
\textbf{Signal} & \textbf{Owner} & \textbf{Meaning} & \textbf{Resolution} \\
\midrule
\textsc{Hold}  & Planner, Builder, & Need user input & Leader forwards question; \\
               & Scriber, Simulator &                 & re-dispatches on answer \\[3pt]
\textsc{Block} & Tester only        & Validation failed & Leader re-dispatches builder \\
               &                    &                   & with failure details \\[3pt]
\textsc{Stop}  & Reviewer only      & Quality gate failed & Leader re-dispatches per \\
               &                    &                     & reviewer's routing \\
\bottomrule
\end{tabular}
\end{table}

Each signal permits a maximum of three retry cycles before escalating to the user, preventing infinite loops while ensuring genuine recovery attempts. \textsc{Block} signals preserve pipeline integrity: the leader re-dispatches the builder with the tester's failure details but never applies fixes directly.

\subsection{Workflow Catalog}
\label{sec:appendix-workflows}

\textsc{StatsClaw} supports ten workflow types, selected by the leader based on the user's natural-language directive. Table~\ref{tab:workflows} lists each type with its agent sequence.

\begin{table}[ht]
\centering
\small
\caption{Workflow types and agent sequences.}\label{tab:workflows}
\begin{tabular}{clp{9cm}}
\toprule
\textbf{\#} & \textbf{Name} & \textbf{Agent Sequence} \\
\midrule
1 & Full Workflow   & leader $\to$ planner $\to$ [builder $\|$ tester $\|$ simulator] $\to$ scriber $\to$ reviewer \\ \hline
\multicolumn{3}{c}{\textit{Simplified Workflows}} \\
2  & Simple Code Fix + Shipping        & leader $\to$ planner $\to$ [builder $\|$ tester] $\to$ scriber $\to$ reviewer $\to$ shipper \\
3  & Documentation          & leader $\to$ planner $\to$ scriber $\to$ reviewer \\
4  & Issue Patrol       & leader scans issues $\to$ per issue: full code workflow \\
5  & Single Issue Fix       & leader $\to$ planner $\to$ [builder $\|$ tester] $\to$ scriber $\to$ reviewer $\to$ shipper \\
6  & Validation         & leader $\to$ tester \\
7  & Code Review        & leader $\to$ reviewer \\
8  & Scheduled Loop     & leader $\to$ recurring inner workflow \\
9 & Extremely Simple Fix         & leader $\to$ builder $\to$ tester \\
10 & Monte Carlo Exercises    & leader $\to$ planner $\to$ [simulator $\|$ tester] $\to$ scriber $\to$ reviewer \\
\bottomrule
\end{tabular}
\end{table}

Brackets with $\|$ indicate parallel dispatch. Workflow selection is semantic: the leader parses the user's directive and matches intent to workflow type. Manual override is available.

\subsection{Runtime Artifacts}
\label{sec:appendix-artifacts}

Every workflow run produces artifacts stored in a run directory within the workspace repository. Table~\ref{tab:artifacts} lists each artifact, its producer, and its purpose.

\begin{table}[ht]
\centering
\small
\caption{Runtime artifacts produced per workflow run.}\label{tab:artifacts}
\begin{tabular}{llp{7cm}}
\toprule
\textbf{Artifact} & \textbf{Producer} & \textbf{Purpose} \\
\midrule
\texttt{request.md}        & Leader    & Scope, acceptance criteria, target repo identity \\
\texttt{impact.md}         & Leader    & Affected files, risk areas, required agents \\
\texttt{status.md}         & Leader    & State machine tracker \\
\texttt{credentials.md}    & Leader    & GitHub access verification \\
\texttt{comprehension.md}  & Planner   & Auditable proof of understanding \\
\texttt{spec.md}           & Planner   & Implementation specification \\
\texttt{test-spec.md}      & Planner   & Test specification \\
\texttt{sim-spec.md}       & Planner   & Simulation specification (workflows 11--12) \\
\texttt{implementation.md} & Builder   & Files changed, design choices \\
\texttt{audit.md}          & Tester    & Per-test results, before/after comparisons \\
\texttt{simulation.md}     & Simulator & DGP, harness design, results \\
\texttt{Architecture.md}   & Scriber   & Mermaid diagrams, reference tables \\
\texttt{log-entry.md}      & Scriber   & Complete process record with handoff notes \\
\texttt{docs.md}           & Scriber   & Documentation change summary \\
\texttt{review.md}         & Reviewer  & Cross-pipeline audit and verdict \\
\texttt{shipper.md}        & Shipper   & Git actions taken, workspace sync status \\
\bottomrule
\end{tabular}
\end{table}

\subsection{Workspace Repository}
\label{sec:appendix-workspace}

The workspace repository is a user-owned GitHub repository, separate from any target codebase, that stores all workflow artifacts and provides cross-session continuity. Its structure is:

\begin{lstlisting}[numbers=none,backgroundcolor=\color{white},frame=none]
workspace/
  <repo-name>/
    context.md            Repo metadata (runtime)
    CHANGELOG.md          Timeline index (pushed)
    HANDOFF.md            Active handoff (pushed)
    docs.md               Latest doc changes (pushed)
    ref/                  Reference materials (pushed)
    runs/
      <request-id>/       Active run artifacts
      <date>-<slug>.md    Completed run logs
    logs/                 Diagnostic logs (local)
    tmp/                  Transient data (local)
\end{lstlisting}

The \texttt{HANDOFF.md} document is updated after every session. Each session's leader reads the previous session's handoff and resumes with full awareness of prior decisions, known issues, and technical insights. This eliminates the resumption cost that typically accompanies interrupted development---particularly valuable in academic settings where development is interleaved with teaching, reviewing, and other obligations.

\subsection{Language Profiles}
\label{sec:appendix-languages}

\textsc{StatsClaw} auto-detects the target language from repository markers and injects language-specific conventions into all agent dispatch prompts. Table~\ref{tab:languages} lists the supported profiles.

\begin{table}[ht]
\centering
\small
\caption{Supported languages and auto-detection markers.}\label{tab:languages}
\begin{tabular}{llll}
\toprule
\textbf{Language} & \textbf{Detection Marker} & \textbf{Ecosystem} & \textbf{Validation} \\
\midrule
R          & \texttt{DESCRIPTION}             & CRAN      & \texttt{R CMD check --as-cran} \\
Python     & \texttt{pyproject.toml}           & PyPI      & \texttt{pytest}, \texttt{tox} \\
TypeScript & \texttt{package.json}             & npm       & \texttt{npm test}, eslint \\
Stata      & \texttt{.ado} files               & SSC       & \texttt{do} file execution \\
Go         & \texttt{go.mod}                   & Go modules & \texttt{go test ./...} \\
Rust       & \texttt{Cargo.toml}               & crates.io & \texttt{cargo test}, clippy \\
C          & \texttt{Makefile} + \texttt{.c}   & System    & Unit test runner \\
C++        & \texttt{CMakeLists.txt} + \texttt{.cpp} & System & Unit test runner \\
\bottomrule
\end{tabular}
\end{table}

Each profile specifies validation commands, documentation conventions (e.g., roxygen2 for R, docstrings for Python), and language-specific builder and tester notes (e.g., numerical stability idioms, strictness levels for test failures).

\section{Probit Specification Document}
\label{sec:probit-spec}

The following pages reproduce the 4-page PDF specification used as source material for the tutorial in Section~\ref{sec:tutorial}.



\section*{Supplementary material}

# Probit Model: Three C++ Estimators via Rcpp

MLE (Newton–Raphson) · Bayesian Gibbs · Metropolis–Hastings

Implementation Specification for StatsClaw Workflow

March 2026

## Contents

## 1 Model and Notation

**Latent variable formulation.** For $i = 1, \ldots, N$:

$$y_i^* = \boldsymbol{x}_i'\boldsymbol{\beta} + \varepsilon_i, \quad \varepsilon_i \sim N(0,1), \quad y_i = \mathbf{1}(y_i^* > 0). \tag{1}$$

This gives $\Pr(y_i = 1 \mid \boldsymbol{x}_i) = \Phi(\boldsymbol{x}_i'\boldsymbol{\beta})$, where $\Phi$ is the standard normal CDF.

**Notation.** $N$ = sample size, $k$ = number of covariates (including intercept), $\boldsymbol{X} \in \mathbb{R}^{N \times k}$ = design matrix (row $i$ is $\boldsymbol{x}_i'$), $\phi(\cdot)$ = standard normal PDF.

**DGP for simulation.** $\Pr(y = 1 \mid x_1) = \Phi(-1 + 0.5\,x_1)$, with $x_1 \sim N(0,1)$. True parameter: $\boldsymbol{\beta}_0 = (-1,\ 0.5)'$.

1

## 2 Method 1: MLE via Newton–Raphson

### 2.1 Log-Likelihood, Gradient, Hessian

$$\ell(\boldsymbol{\beta}) = \sum_{i=1}^{N} \Big[ y_i \log \Phi(\boldsymbol{x}_i'\boldsymbol{\beta}) + (1 - y_i) \log \big(1 - \Phi(\boldsymbol{x}_i'\boldsymbol{\beta})\big) \Big] \quad (2)$$

Define $q_i = 2y_i - 1 \in \{-1, +1\}$, $\eta_i = \boldsymbol{x}_i'\boldsymbol{\beta}$, and the inverse Mills ratio $\lambda_i = \phi(q_i \eta_i)/\Phi(q_i \eta_i)$.

$$\text{Gradient:} \quad \nabla \ell = \boldsymbol{X}'\boldsymbol{w}, \qquad w_i = q_i \lambda_i \quad (3)$$

$$\text{Hessian:} \quad \nabla^2 \ell = -\boldsymbol{X}'\boldsymbol{D}\boldsymbol{X}, \qquad d_i = \lambda_i(\lambda_i + q_i \eta_i) > 0 \quad (4)$$

### 2.2 Algorithm

**Algorithm 1** Probit MLE

1: Initialize $\boldsymbol{\beta}^{(0)} = \boldsymbol{0}$
2: **for** $t = 1, \ldots, T_{\max}$ **do**
3:     Compute $\boldsymbol{w}, \boldsymbol{D}$ from current $\boldsymbol{\beta}$
4:     $\boldsymbol{\beta}^{(t)} \leftarrow \boldsymbol{\beta}^{(t-1)} - (\nabla^2 \ell)^{-1} \nabla \ell$      $\triangleright = \boldsymbol{\beta}^{(t-1)} + (\boldsymbol{X}'\boldsymbol{D}\boldsymbol{X})^{-1}\boldsymbol{X}'\boldsymbol{w}$
5:     Stop if $\|\boldsymbol{\beta}^{(t)} - \boldsymbol{\beta}^{(t-1)}\| < 10^{-8}$
6: **end for**
7: **Output:** $\hat{\boldsymbol{\beta}}$, $\text{SE}_j = \sqrt{[(\boldsymbol{X}'\boldsymbol{D}\boldsymbol{X})^{-1}]_{jj}}$, $\ell(\hat{\boldsymbol{\beta}})$

### 2.3 C++ Interface

```
// [[Rcpp::export]]
Rcpp::List probit_mle(const arma::mat& X, const arma::vec& y,
                      int max_iter = 100, double tol = 1e-8);
// Returns: coefficients (vec), vcov (mat), se (vec), loglik (double)
```

**Numerical stability:** Use `R::pnorm(x, 0, 1, 1, 1)` for $\log \Phi(x)$. Solve $\boldsymbol{H}\boldsymbol{s} = \boldsymbol{g}$ via `arma::solve`, not explicit inversion.

## 3 Method 2: Bayesian Gibbs Sampler (Albert–Chib)

### 3.1 Prior and Conditional Posteriors

Prior: $\boldsymbol{\beta} \sim N(\boldsymbol{\beta}_0, \boldsymbol{\Sigma}_0)$. Default diffuse: $\boldsymbol{\beta}_0 = \boldsymbol{0}$, $\boldsymbol{\Sigma}_0 = 100\boldsymbol{I}$.

Introduce latent $\boldsymbol{y}^* = (y_1^*, \ldots, y_N^*)'$ where $y_i^* \sim N(\boldsymbol{x}_i'\boldsymbol{\beta}, 1)$. The Gibbs sampler alternates:

**Step 1 — Sample $\boldsymbol{\beta}$** (closed-form normal):

$$\boldsymbol{\beta} \mid \boldsymbol{X}, \boldsymbol{y}^* \sim N(\tilde{\boldsymbol{\beta}}, \tilde{\boldsymbol{\Sigma}}) \quad (5)$$

$$\tilde{\boldsymbol{\Sigma}} = \big(\boldsymbol{\Sigma}_0^{-1} + \boldsymbol{X}'\boldsymbol{X}\big)^{-1}, \qquad \tilde{\boldsymbol{\beta}} = \tilde{\boldsymbol{\Sigma}}\big(\boldsymbol{\Sigma}_0^{-1}\boldsymbol{\beta}_0 + \boldsymbol{X}'\boldsymbol{y}^*\big) \quad (6)$$

*Derivation.* The likelihood of $\boldsymbol{y}^*$ given $\boldsymbol{\beta}$ is $p(\boldsymbol{y}^* \mid \boldsymbol{\beta}) \propto \exp\big(-\tfrac{1}{2}(\boldsymbol{y}^* - \boldsymbol{X}\boldsymbol{\beta})'(\boldsymbol{y}^* - \boldsymbol{X}\boldsymbol{\beta})\big)$. Combining with the prior and completing the square in $\boldsymbol{\beta}$ gives the normal posterior above. The precision matrix is $\boldsymbol{\Sigma}_0^{-1} + \boldsymbol{X}'\boldsymbol{X}$ and the precision-weighted mean is $\boldsymbol{\Sigma}_0^{-1}\boldsymbol{\beta}_0 + \boldsymbol{X}'\boldsymbol{y}^*$.

**Step 2 — Sample each $y_i^*$** (truncated normal):

$$y_i^* \mid (y_i = 1, \boldsymbol{\beta}) \sim \text{TN}(\boldsymbol{x}_i'\boldsymbol{\beta}, 1; 0, +\infty), \qquad y_i^* \mid (y_i = 0, \boldsymbol{\beta}) \sim \text{TN}(\boldsymbol{x}_i'\boldsymbol{\beta}, 1; -\infty, 0) \qquad (7)$$

*Derivation.* Since $y_i^* \sim N(\boldsymbol{x}_i'\boldsymbol{\beta}, 1)$ and $y_i = \mathbf{1}(y_i^* > 0)$, the conditional $y_i^* \mid y_i$ is the original normal truncated to the region consistent with the observed $y_i$.

**Truncated normal sampling** (inverse CDF): draw $u \sim \text{Unif}(0,1)$ and compute

$$y_i^* = \mu_i + \Phi^{-1}\big(\Phi(a - \mu_i) + u\left[\Phi(b - \mu_i) - \Phi(a - \mu_i)\right]\big), \qquad (8)$$

where $\mu_i = \boldsymbol{x}_i'\boldsymbol{\beta}$ and $(a, b)$ is $(0, +\infty)$ or $(-\infty, 0)$.

### 3.2 Algorithm

---
**Algorithm 2** Probit Gibbs Sampler
---
1: Precompute $\tilde{\boldsymbol{\Sigma}} = (\boldsymbol{\Sigma}_0^{-1} + \boldsymbol{X}'\boldsymbol{X})^{-1}$ and its Cholesky $\boldsymbol{L}$ ▷ constant, compute once
2: Initialize $\boldsymbol{\beta}^{(0)} = \mathbf{0}$
3: **for** $t = 1, \ldots, T$ **do**
4:     **for** $i = 1, \ldots, N$ **do** ▷ Step 2: sample latent data
5:         $y_i^{*(t)} \sim \text{TN}(\boldsymbol{x}_i'\boldsymbol{\beta}^{(t-1)}, 1;\ \text{bounds from } y_i)$
6:     **end for**
7:     $\tilde{\boldsymbol{\beta}} \leftarrow \tilde{\boldsymbol{\Sigma}}(\boldsymbol{\Sigma}_0^{-1}\boldsymbol{\beta}_0 + \boldsymbol{X}'\boldsymbol{y}^{*(t)})$ ▷ Step 1: sample $\boldsymbol{\beta}$
8:     $\boldsymbol{\beta}^{(t)} \leftarrow \tilde{\boldsymbol{\beta}} + \boldsymbol{L}\boldsymbol{z},\ \boldsymbol{z} \sim N(\mathbf{0}, \boldsymbol{I}_k)$
9: **end for**
10: Discard first $B$ draws (burn-in). **Output:** posterior draws $\{\boldsymbol{\beta}^{(t)}\}_{t=B+1}^T$
---

### 3.3 C++ Interface

```
// [[Rcpp::export]]
Rcpp::List probit_gibbs(const arma::mat& X, const arma::vec& y,
                        int n_iter = 3500, int burn_in = 500,
                        Nullable<arma::vec> beta0, Nullable<arma::mat> Sigma0);
// Returns: beta_draws (mat), posterior_mean (vec), posterior_sd (vec)
```

**Key optimization:** $\tilde{\boldsymbol{\Sigma}}$ and $\boldsymbol{L}$ depend only on $\boldsymbol{X}$ and the prior — precompute once outside the loop.

## 4 Method 3: Metropolis–Hastings MCMC

### 4.1 Log-Posterior and Proposal

Log-posterior (up to constant):

$$\log p(\boldsymbol{\beta} \mid \boldsymbol{X}, \boldsymbol{y}) = \underbrace{\sum_i \left[ y_i \log \Phi(\eta_i) + (1 - y_i) \log(1 - \Phi(\eta_i)) \right]}_{\ell(\boldsymbol{\beta})} - \tfrac{1}{2}(\boldsymbol{\beta} - \boldsymbol{\beta}_0)'\boldsymbol{\Sigma}_0^{-1}(\boldsymbol{\beta} - \boldsymbol{\beta}_0) \qquad (9)$$

Random walk proposal: $\boldsymbol{\beta}' = \boldsymbol{\beta}^{(t)} + \boldsymbol{\varepsilon},\ \ \boldsymbol{\varepsilon} \sim N\big(\mathbf{0},\ s^2\,(\boldsymbol{X}'\boldsymbol{X})^{-1}\big)$.
Accept with probability $\alpha = \min\big(1,\ \exp[\log p(\boldsymbol{\beta}' \mid \cdot) - \log p(\boldsymbol{\beta}^{(t)} \mid \cdot)]\big)$.

## 4.2 Algorithm

---
**Algorithm 3** Probit Metropolis–Hastings
---
1: Initialize $\boldsymbol{\beta}^{(0)}$ from MLE. Precompute proposal Cholesky $\boldsymbol{L}_C = \text{chol}(s^2(\boldsymbol{X}'\boldsymbol{X})^{-1})$.
2: **for** $t = 1, \ldots, T$ **do**
3:     Propose: $\boldsymbol{\beta}' = \boldsymbol{\beta}^{(t-1)} + \boldsymbol{L}_C \boldsymbol{z}$, $\boldsymbol{z} \sim N(\boldsymbol{0}, \boldsymbol{I}_k)$
4:     $\log r = \log p(\boldsymbol{\beta}' \mid \cdot) - \log p(\boldsymbol{\beta}^{(t-1)} \mid \cdot)$
5:     If $\log u < \log r$ ($u \sim \text{Unif}$): accept $\boldsymbol{\beta}^{(t)} = \boldsymbol{\beta}'$; else: $\boldsymbol{\beta}^{(t)} = \boldsymbol{\beta}^{(t-1)}$
6: **end for**
7: Discard burn-in. **Output:** posterior draws, acceptance rate (target: 25–40%)
---

## 4.3 C++ Interface

```
// [[Rcpp::export]]
Rcpp::List probit_mh(const arma::mat& X, const arma::vec& y,
                     int n_iter = 10000, int burn_in = 2000,
                     double scale = 1.0,
                     Nullable<arma::vec> beta0, Nullable<arma::mat> Sigma0,
                     Nullable<arma::vec> init);
// Returns: beta_draws (mat), posterior_mean (vec), posterior_sd (vec),
//          acceptance_rate (double)
```

**Warm start:** Initialize from MLE for faster convergence. Tune $s$ so acceptance is 25–40%.

## 5 Shared Utilities (`utils.cpp`)

```
double safe_log_Phi(double x);    // R::pnorm(x, 0, 1, 1, 1) - log-space
double safe_Phi(double x);        // R::pnorm(x, 0, 1, 1, 0)
double rtruncnorm(double mu,      // Inverse-CDF truncated normal sampler
                  double lo, double hi);
```

## 6 Monte Carlo Simulation Study

### 6.1 Design

| | |
|---|---|
| DGP | $\Pr(y = 1 \mid x_1) = \Phi(-1 + 0.5\, x_1)$, $x_1 \sim N(0,1)$ |
| True parameters | $\boldsymbol{\beta}_0 = (-1,\ 0.5)'$ |
| Sample sizes $N$ | $\{200, 500, 1000, 5000\}$ |
| Replications | $R = 500$ per scenario |
| MLE | max 100 iter, tol $10^{-8}$ |
| Gibbs | 3500 iter, burn-in 500 |
| MH | 10000 iter, burn-in 2000, $s = 1$ |
| Prior (Gibbs & MH) | $\boldsymbol{\beta}_0 = \boldsymbol{0}$, $\boldsymbol{\Sigma}_0 = 100 \boldsymbol{I}_2$ |

4

## 6.2 Metrics (per method, per $\beta_j$, across $R$ replications)

$$\text{Bias}_j = \frac{1}{R}\sum_{r=1}^{R}(\hat{\beta}_j^{(r)} - \beta_{0,j}) \qquad \text{RMSE}_j = \sqrt{\frac{1}{R}\sum_{r=1}^{R}(\hat{\beta}_j^{(r)} - \beta_{0,j})^2} \qquad (10)$$

$$\text{Coverage}_j = \frac{1}{R}\sum_{r=1}^{R}\mathbf{1}(\beta_{0,j} \in \text{CI}_{95\%}^{(r)}) \qquad \text{Time} = \frac{1}{R}\sum_{r=1}^{R}t^{(r)} \text{ (seconds)} \qquad (11)$$

For MLE: CI from asymptotic normality $\hat{\beta}_j \pm 1.96 \cdot \text{SE}_j$. For Gibbs/MH: CI from 2.5% and 97.5% quantiles of posterior draws.

## 6.3 Acceptance Criteria

**C1** MLE coefficients match `glm(..., family=binomial(link="probit"))` within 0.05

**C2** Gibbs and MH posterior means within 0.1 of MLE for $N \geq 500$

**C3** MH acceptance rate between 20% and 50%

**C4** 95% CI coverage between 90% and 99% for all methods

**C5** MLE at least 5× faster than Gibbs (C++ vs C++)

**C6** `R CMD check` passes with 0 errors, 0 warnings

**C7** Package installs via `R CMD INSTALL` and all functions callable from R

5



\end{document}